\documentclass[aps,prd,twocolumn,groupedaddress,amsmath,amssymb,reprint]{revtex4-1}

\pdfoutput=1

\usepackage{graphicx}  
\usepackage{dcolumn}   
\usepackage{bm}        
\usepackage{verbatim}   
\usepackage{multirow}
\usepackage{xcolor}

\usepackage{hyperref}
\hypersetup{colorlinks=true}

\newcommand{\apjl}{Astrophys.~J.~Lett.}
\newcommand{\mnras}{Mon.~Not.~R.~Astron.~Soc.}
\newcommand{\aap}{Astron.~Astrophys.}
\newcommand{\aj}{Astron.~J.}
\newcommand{\physrep}{Phys.~Rep.}
\newcommand{\araa}{Ann.~Rev.~Astron.~Astrophys.}

\newcommand{\jcap}{J.~Cosmol.~Astropart.~Phys.}

\newcommand{\ssr}{Space Sci. Rev.}

\begin{document}

\title{Effects of Self-Calibration of Intrinsic Alignment on Cosmological Parameter Constraints from Future Cosmic Shear Surveys}
\author{Ji Yao}
\email{jxy131230@utdallas.edu}
\affiliation{Department of Physics, The University of Texas at Dallas, Dallas, TX 75080}
\author{Mustapha Ishak}
\email{mxi054000@utdallas.edu}
\affiliation{Department of Physics, The University of Texas at Dallas, Dallas, TX 75080}
\author{Weikang Lin}
\affiliation{Department of Physics, The University of Texas at Dallas, Dallas, TX 75080}
\author{Michael Troxel}
\affiliation{Department of Physics, Ohio State University, Columbus, OH 43210}
\date{\today}
\begin{abstract}
Intrinsic alignments (IA) of galaxies have been recognized as one of the most serious contaminants to weak lensing. These systematics need to be isolated and mitigated in order for ongoing and future lensing surveys to reach their full potential.
The IA self-calibration (SC) method was shown in previous studies to be able to reduce the GI contamination by up to a factor of 10 for the 2-point and 3-point correlations.
The SC method does not require the assumption of an IA model in its working and can extract the GI signal from the same photo-z survey offering the possibility to test and understand structure formation scenarios and their relationship to IA models.
In this paper, we study the effects of the IA SC mitigation method on the precision and accuracy of cosmological parameter constraints from future cosmic shear surveys LSST, WFIRST and Euclid.
We perform analytical and numerical calculations to estimate the loss of precision and the residual bias in the best fit cosmological parameters after the self-calibration is performed. We take into account uncertainties from  photometric redshifts and the galaxy bias.
We find that the confidence contours are slightly inflated from applying the SC method itself while a significant increase is due to the inclusion of the photo-z uncertainties. The bias of cosmological parameters is reduced from several-$\sigma$, when IA is not corrected for, to below 1-$\sigma$ after SC is applied.
These numbers are comparable to those resulting from applying the method of marginalizing over IA model parameters despite the fact that the two methods operate very differently.
We conclude that implementing the SC for these future cosmic-shear surveys will not only allow one to efficiently mitigate the GI contaminant but also help to understand their modeling and link to structure formation.
 \end{abstract}

\maketitle

\section{Introduction}

Weak gravitational lensing is one of the primary cosmological probes used to study the properties and distribution of matter in the universe as well as to constrain the nature of the dark energy associated with cosmic acceleration, see for example  \cite{Kaiser1992,Hu1999,Heavens2000,Bacon2001,Ishak2005,Huterer2002,Ishak2007,Joudaki2009,Weinberg2013,Kilbinger2015}. By constraining the growth rate of structures in the universe, weak lensing also provides a means to test gravity theories at cosmological scales, see for example  \cite{Ishak2006,Linder2007,2011PhRvD..84b3012D,Heavens2009,Dossett2012,Alam2017,Dossett2013}. Some completed or ongoing surveys have already provided some useful constraints on cosmological parameters \cite{CFHTLenSHeymans,Hildebrandt2016,DES2016,DLS2016}. Finally, weak lensing is complementary and orthogonal to other cosmological probes and thus provides a means to break degeneracies in the parameter space \cite{Tereno2005,Ishak2005,Munshi2008,DLS2016,Semboloni2011,DossettEtAl2015} and fulfill the endeavor of precision cosmology.

However, weak lensing is not immune to serious systematic effects \cite{BridleKing,Bacon2001,Bernstein2002,Erben2001,Heymans2004,Hirata2003,Ishak2004,Faltenbacher2009}.
At the forefront of these systematics are the intrinsic  alignment (IA) of galaxies that produce correlations that contaminate the lensing signal. There are two IA types that interfere with lensing. The first one, called the II signal, is due to the intrinsic ellipticity correlations that exist between close galaxies formed in the same tidal gravitational field. The second one, called the GI signal, is due to a \textcolor{black}{correlation/anti-correlation} between a galaxy tangentially lensed by a foreground structure and a galaxy near this lens structure oriented radially towards it \cite{Catelan2001,Hirata2003,King2005,Mandelbaum2006,Hirata2007}. In addition to these 2-point IA correlations, there are also their 3-point correlation analogues noted as III, GGI and GII \cite{Semboloni2009,Troxel2012,Refregier2003}. It was shown in \cite{Hirata2004,BridleKing} that when weak lensing alone is used and IA is ignored, IA can bias the amplitude of the matter power spectrum by up to $30\%$ \cite{Hirata2004} and the equation of state of dark energy by up to $50\%$ \cite{BridleKing}. We refer the reader to recent reviews and references therein on the subject of intrinsic alignments \cite{Bernstein2009,TroxelIshak,Joachimi2013,Kilbinger2015,Krause2016,Blazek2015}.


These systematics need to be mitigated in order for weak lensing to reach its full potential. While the II and III correlations of IA can be, in principle, greatly reduced with photo-z by using cross-spectra of galaxies in two different redshift bins \cite{SC.Zhang,Refregier2003}, so that the galaxies are separated by large enough distances to ensure that the tidal effect is weak, this does not work for the GI, GGI, and GII types which happen between galaxies at different redshifts and large separations.
Consequently, other methods have been proposed to mitigate the GI signal. One of them is the usual marginalization method where the IA signal is introduced using a given model with parameters to be constrained along with the cosmological ones \cite{BridleKing,Krause2016,Krause2016b,Joachimi2013,Blazek2015,Joachimi2015}. A second method, called the self-calibration, operates differently and aims at extracting the GI signal itself from the same cosmic shear survey. As we detail in the next section, the self-calibration technique can be applied to photometric surveys and the GI signal is dhetermined from the shear-galaxy density correlation taking into account the geometry, the photo-z and the galaxy bias. These works \cite{SC.Zhang,TroxelIshak,Zhang2010,Troxel2012,Troxel2012b,Troxel2012c} \textcolor{black}{showed that with a total extra systematic error of $10\%$ \cite{SC.Zhang} in the SC scaling relation}, the method is able to reduce the IA contamination by a factor of 10 without throwing away too much of the lensing signal. Ref. \cite{Troxel2012,Troxel2012b,Troxel2012c,TroxelIshak} extended the SC formalism to the 3-point correlations involving the GGI and GII signals as well using the CMB lensing \cite{TroxelIshak,Troxel2014}. The self-calibration method has also been extended to mitigate the cross-correlation between cosmic microwave background lensing and galaxy IA as a contaminant to the gravitational lensing cross-correlated probes \cite{Troxel2014}.

We consider the two methods above, i.e. marginalization over the IA model and the self-calibration as complementary approaches that can serve to cross-validate significant results from weak lensing surveys and mitigation of IA signals. It is worth noting that the self-calibration does not require one to assume a GI model in its process and allows to for the extraction of the signal from a cosmic shear survey. This provides opportunities to study the IA signal and information on structure formation. Indeed, studying IA has two benefits: the first is to clean the weak lensing signals; and the second is to use them to investigate and understand better structure formation scenarios \cite{Kilbinger2015,Joachimi2013,TroxelIshak}.

In this paper, we develop a forecast formalism for the self-calibration in order to assess its effects and the accuracy and precision of cosmological parameter estimation using future photometric surveys. We apply the formalism with a special focus on the LSST survey but also extend the analysis to WFIRST and Euclid. We also derive the corresponding effects from using the marginalization method and compare the results to those of the self-calibration.

The paper is organized as follows. In Section \ref{Section SC} we describe the 2-point Self Calibration technique of IA. We derive the Fisher formalism for the self-calibration and cosmological parameters forecasts in Section \ref{Section Fisher}. The details for the models of photo-z, galaxy bias and IA are included in Section \ref{Section models}. The results are presented in Section \ref{Section results}. A discussion and summary are given in Section \ref{Section discussion}.

\section{The Self Calibration Technique} \label{Section SC}

\subsection{Method}

The measured shear contains $\gamma^G+\gamma^I+\gamma^N$, with $G$ being the gravitational shear, $I$ the intrinsic alignment, and \textcolor{black}{$N$ the measurement noise and shape noise}. Thus the observed contaminated angular cross correlation power spectra are \cite{BridleKing}:
\begin{subequations}
\begin{align} \label{observables}
C^{(1)}_{ij}(l)&=C^{GG}_{ij}(l)+C^{IG}_{ij}(l)+C^{GI}_{ij}(l)+C^{II}_{ij}(l)+\delta_{ij}C^{GG,N}_{ii},\\
C^{(2)}_{ii}(l)&=C^{gG}_{ii}(l)+C^{gI}_{ii}(l), \label{C2}\\
C^{(3)}_{ii}(l)&=C^{gg}_{ii}(l)+\delta_{ij}C^{gg,N}_{ii}. \label{C3}
\end{align}
\end{subequations}

\begin{figure*}
\includegraphics[width=2.0\columnwidth]{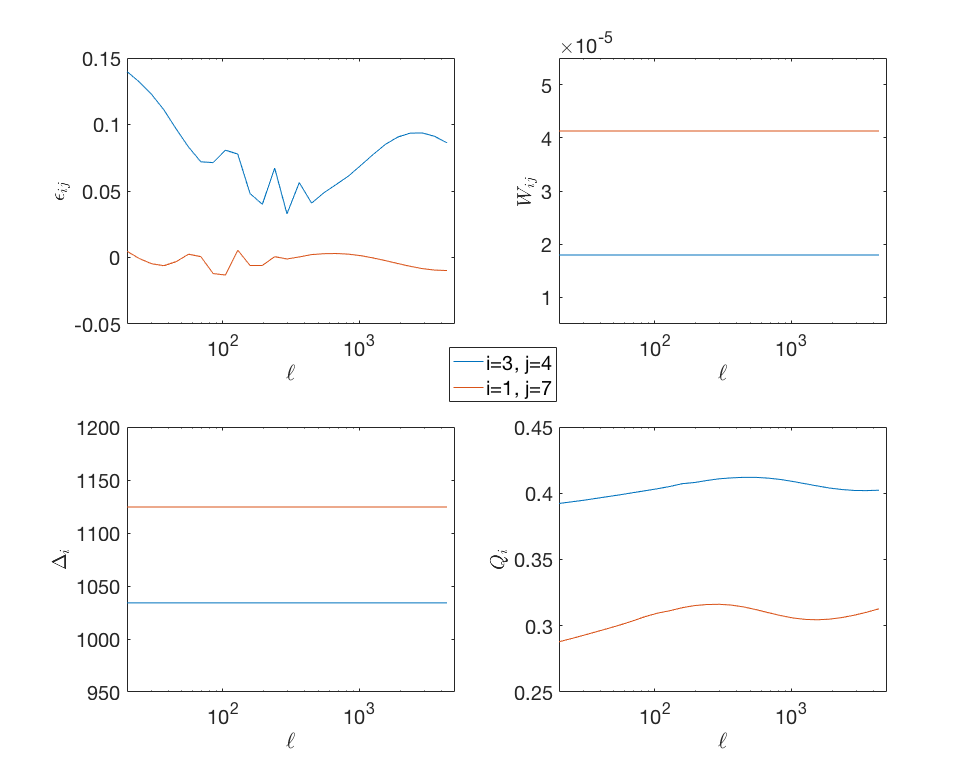}
\caption{Numerical calculations for various quantities used in SC. Top-left panel shows the fractional residual bias of SC, with the definition $\epsilon_{ij}\equiv (C^{IG(true)}_{ij}-C^{IG(SC)}_{ij})/{C^{IG(SC)}_{ij}}$ as in Eq.\,\eqref{epsilon}. The comparison between blue and red gives an example that in close bins, the efficiency of SC is lower (residual error higher) than bins that are far away. Note that $\epsilon_{ij}\lesssim 0.1$ for all bin-pairs and is $\sim 0$ for far-separated bins. The other 3 panels are the theoretical quantities (i.e. $W_{ij}$, $\Delta_i$ and $Q_i$, which are calculated by Eq.\,\eqref{Wij}, \eqref{Delta_i} and \eqref{Q}) and used by SC, through Eq.\,\eqref{scaling-1} and \eqref{scaling-2}.
\label{Reproduce Zhang}}
\end{figure*}

In Eq.~(\ref{observables}), the shear spectrum $C^{GG}$ is what is used in Weak Lensing (WL) to constrain cosmological models. We can use the Self-Calibration technique to subtract the major contamination term $C^{IG}_{ij}$ as well as minimize the effect of $C^{GI}_{ij}$ and $C^{II}_{ij}$ by its particular binning \cite{SC.Zhang}. Based on the difference between gravitational shear and intrinsic shear, this technique uses a scaling relation (the approximation Eq.~(\ref{scaling-1}) that will be discussed further) to obtain $C^{IG}_{ij}$ from other photo-z observables. It doesn't require us to have any a priori knowledge about the intrinsic alignment model itself.

In Eq.\,\eqref{C2} and \eqref{C3}, \textcolor{black}{$C^{(2)}$ is the galaxy-galaxy lensing spectrum, with the contamination from the intrinsic alignment, and $C^{(3)}$ is the galaxy-galaxy clustering spectrum.} $C^{gI}$ is the galaxy-intrinsic spectrum, $C^{gG}$ is the galaxy-shear spectrum, and $C^{gg}$ is the galaxy-galaxy spectrum. N denotes the noise spectrum. It is required that we use the same clustering sample of galaxies as the lensing sample for the scaling relation Eq.\,\eqref{scaling-1} of SC to work.

\textcolor{black}{However, even if Eq.\,\eqref{C2} and \eqref{C3} are used in deriving the SC method, we note that they are only used to get the information of IA ( i.e. the $C^{Ig}$ spectrum) in oder to derive the scaling relation Eq.\,\eqref{scaling-2}), while the constraining power of galaxy-galaxy lensing and galaxy-galaxy clustering are not used in either the SC analysis (in Subsection \ref{subsection using SC})  or the IA-marginalization analysis (in Subsection \ref{subsection marginalization}). \textcolor{black}{In other words, SC uses the galaxy-galaxy lensing and clustering infomation to treat IA, but not any additional cosmological infomation they contain.} In this work we focus on the Self-Calibration of the IA problem in cosmic shear. In future work, we will explore the role of SC in the joint analysis of cosmic shear, galaxy-galaxy lensing and galaxy-galaxy clustering \cite{Joudaki2017,Krause2017}.}

The useful difference is that $\gamma^I$ correlates with galaxies at both higher redshift $z\ge z^I$ and lower redshift $z\le z^I$, while $\gamma^G$ only correlates with galaxies at lower redshift. Note here $z$ is the true redshift. In practice, because of survey limits, we normally use photometric redshift in order to obtain more data, and this relation of the difference between correlations doesn't always hold for photo-z. Instead, we separate the survey into several photo-z bins.

For redshift bins $i<j$, which means the photo-z obeys $z^P_i<z^P_j$, $C^{GI}_{ij}(l) \ll C^{IG}_{ij}(l)$ holds for reasonably small catastrophic error \cite{SC.Zhang}. And $C^{II}$ only exists at small line-of-sight separation, which means $C^{II}_{ij}(l)$ is also negligible. Thus with this binning method, $C^{GI}_{ij}$ and $C^{II}_{ij}$ are automatically minimized, so we only need to worry about the major contamination term $C^{IG}_{ij}(l)$ in Eq.\,\eqref{observables}. The SC presented here relies on two relations \cite{SC.Zhang} that we will discuss as follows.

The first scaling relation is derived from rewriting the Limber integration of $C^{IG}$ and $C^{Ig}$:
\begin{subequations}
	\begin{align}
		C^{IG}_{ij}(\ell)&=\int_{0}^{\infty}\frac{n_i(\chi)q_j(\chi)}{\chi^2}P_{\delta,\gamma^I}(k;\chi)d\chi \label{IGij_Limber},\\
		C^{Ig}_{ii}(\ell)&=\int_{0}^{\infty}\frac{n_i^2(\chi)b_g(k,z)}{\chi^2}P_{\delta,\gamma^I}(k;\chi)d\chi \label{Igii_Limber},
	\end{align}
\end{subequations}
where $n_i(\chi)$ is the galaxy distribution in the i-th redshift bin, $q_j(\chi)$ is the lensing window function for the j-th redshift bin, $b_g(k,z)$ is the galaxy bias, and $P_{\delta,\gamma^I}(k;\chi)$ is the 3-D matter-intrinsic cross correlation power spectrum. Under small-bin approximation (normally $\Delta z\leq 0.2$ \cite{SC.Zhang}), components in the the above two integrations (Eq.\,\eqref{IGij_Limber} and \eqref{Igii_Limber}) change slowly so that we have the following approximation:
\begin{subequations}
	\begin{align}
		C^{IG}_{ij}(\ell)&\approx \frac{W_{ij}}{\chi_i^2}P_{\delta,\gamma^I}(k_i=\frac{\ell}{\chi_i};\chi_i) \label{IG limber},\\
		C^{Ig}_{ii}(\ell)&\approx b_i(\ell)\frac{1}{\chi_i^2\Delta_i}P_{\delta,\gamma^I}(k_i=\frac{\ell}{\chi_i};\chi_i) , \label{Ig limber}
	\end{align}
\end{subequations}
where
\begin{subequations}
\begin{align}
W_{ij}&\equiv \int_{0}^{\infty}dz_L\int_{0}^{\infty}dz_S
[W_L(z_L,z_S)n_i(z_L)n_j(z_S)],\label{Wij} \\
\Delta_i^{-1}&\equiv \int_{0}^{\infty}n_i^2(z)\frac{dz}{d\chi}dz, \label{Delta_i}
\end{align}
\end{subequations}
which are shown in Fig.\,\ref{Reproduce Zhang} (top-right panel and bottom-left panel). 

Therefore $C^{IG}_{ij}$ and $C^{Ig}_{ii}$ can be connected without concerning the 3-D spectrum $P_{\delta,\gamma^I}(k;\chi)$, and consequently the IA model. In Eq.~(\ref{Wij}), $W_L$ is the lensing kernel:
\begin{equation}
W_L(z_L,z_S)=\begin{cases}
\frac{3}{2}\Omega_m\frac{H_0^2}{c^2}(1+z_L)\chi_L(1-\frac{\chi_L}{\chi_S}) &\text{for $z_L<z_S$}\\ 0 &\text{otherwise}
\end{cases}.
\end{equation}

Based on the above approximations, the first scaling relation is obtained:
\begin{equation} \label{scaling-1}
C^{IG}_{ij}(l)\simeq \frac{W_{ij}(l)\Delta_i(l)}{b_i(l)}C^{Ig}_{ii}(l).
\end{equation}

In Eq.\,\eqref{Ig limber} and \eqref{scaling-1}, $b_i(l)\approx\int_0^\infty b_g(k,z)n_i(z)dz$ is the averaged galaxy bias in each redshift bin, where $b_g$ is the galaxy bias given by $b_g(k=l/\chi,z)=\delta_g/\delta$. (This approximation is used for this theoretical forecast. In practice, the averaged bias $b_i$ is calculated using another approximation $C^{gg}\approx b_i^2C^{mm}$, where $C^{mm}$ is the matter power spectrum.) The $n_i(z)$ in Eqs\,\eqref{Wij} and \eqref{Delta_i} are the normalized true-z distribution within the i-th tomographic bin. The details about the galaxy bias and photo-z being used in this work will be discussed in Section \ref{Section models}.

This scaling relation [Eq.~(\ref{scaling-1})] hold for reasonably small photo-z bin width, \textcolor{black}{while the terms inside the Limber integration of Eq.\,\eqref{IG limber} and \eqref{Ig limber} change slowly with time. Because of this, the scaling relation Eq.\,\eqref{scaling-1} holds for any IA model or linear galaxy bias model as long as they are not very sensitive to the changes in redshift $z$. For the non-linear galaxy bias model, the scaling relation Eq.\,\eqref{scaling-1} needs to be adjusted like in Ref.\,\cite{Troxel2012}, otherwise there will be a drop in its efficiency. In this work we use the presented scaling relation of Eq.\,\eqref{scaling-1} and leave further discussion on non-linear bias to the future, as linear galaxy biasing has also been required in recent combined probes works \cite{Joudaki2017,Krause2017}. The impact of stochastic galaxy bias on SC has been discussed in Ref.\,\cite{SC.Zhang} and the impact of non-linear galaxy bias on the 3-point SC has been discussed in Ref.\,\cite{Troxel2012}. The effect of catastrophic photo-z outliers on SC (on both Eq.\,\eqref{scaling-1} and Eq.\,\eqref{Q} below) has also been studied in Ref.\,\cite{SC.Zhang} and is expected to be negligible.}


The second relation is based on the difference between gravitational shear and intrinsic alignment. Photo-z measurement contains useful information that allow us to separate $C^{Gg}_{ii}$ and $C^{Ig}_{ii}$:

The I-g correlation does not depend on the ordering along the line-of-sight. Pairs with $z^P_G>z^P_g$ are statistically identical to those with $z^P_G<z^P_g$ within the same redshift bin.

The G-g correlation, on the other hand, does depend on the ordering. Pairs with $z^P_G>z^P_g$ have stronger correlation than those with $z^P_G<z^P_g$.

The SC method introduces another observable $C^{(2)}_{ii}|_S(l)$ in which ``S'' stands for only correlating pairs with $z^P_G<z^P_g$. Apparently $C^{Ig}_{ii}|_S(l)=C^{Ig}_{ii}(l)$, while $C^{Gg}_{ii}|_S(l)<C^{Gg}_{ii}(l)$. Define:
\begin{equation} \label{Q}
Q_i(l)\equiv \frac{C^{Gg}_{ii}|_S(l)}{C^{Gg}_{ii}(l)}.
\end{equation}
Usually $0<Q<1$. $Q=1$ if photo-z's are completely wrong, and $Q=0$ when photo-z's are perfect. The bottom-right panel in Fig.\,\ref{Reproduce Zhang} gives an example of the values of $Q_i$. Similar expressions have been used in Ref\,\cite{HirataMandelbaum2004,Blazek2012}. Then the two observables read:

\begin{subequations}
\begin{align}
C^{(2)}_{ii}(l)&=C^{Ig}_{ii}(l)+C^{Gg}_{ii}(l),\\
C^{(2)}_{ii}|_S(l)&=C^{Ig}_{ii}(l)+C^{Gg}_{ii}|_S(l).
\end{align}
\end{subequations}

The above two equations provide us the second relation:
\begin{equation} \label{scaling-2}
C^{Ig}_{ii}(l)=\frac{C^{(2)}_{ii}|_S(l)-Q_i(l)C^{(2)}_{ii}(l)}{1-Q_i(l)},
\end{equation}
gives $C^{Ig}$ in terms of the SC-observables and photo-z properties. \textcolor{black}{The measurement of SC is based on the detection of Eq.\,\eqref{scaling-2}. The detectability and S/N are discussed in section 3 of Ref.\,\cite{SC.Zhang}. In this work we estimated the value of $|C^{Ig}|/\Delta C^{Ig}$ (uncertainty calculated by Eq.\,\eqref{shot error 1}) as a way to show the detectability, higher value meaning more detectable. The value is generally greater than 1 for a LSST-like survey with $\ell$ range $20<\ell<5000$. At low $\ell$, for example $\ell<50$, the lowest value can be less than (but close to) 1 for some redshift bins. For high $\ell$, the value is generally greater than 10.}

From the above discussion, it has been shown that the SC method is IA-model independent, as long as the IA signal changes slowly within each redshift bin. The small-bin approximation used in SC is not a strong assumption, and it is likely to be satisfied for the future photometric survey. In this paper we will take one common IA model as an example to show the performance of SC. We however note that our framework, especially Eq.\,\eqref{scaling-1}, holds only for a linear bias model. We assume the nonlinear term in the bias model is subdominant. Note linear biasing has also been required in recent combined probes studies \cite{Joudaki2017,Krause2017}.  For a nonlinear bias model, SC can be performed with a modification of the framework, see Ref.\,\cite{Troxel2012}. \textcolor{black}{The study of the impact of non-linear bias on SC will thus require a modified form of the scaling relation and a different error analysis than what has been shown in Fig.\,\ref{statistical error}. This should be explored in future works and is beyond the scope of this paper.} We will discuss the performance of SC in the following subsection.

\subsection{SC Performance}

\begin{figure}[t!]
\includegraphics[width=0.9\columnwidth]{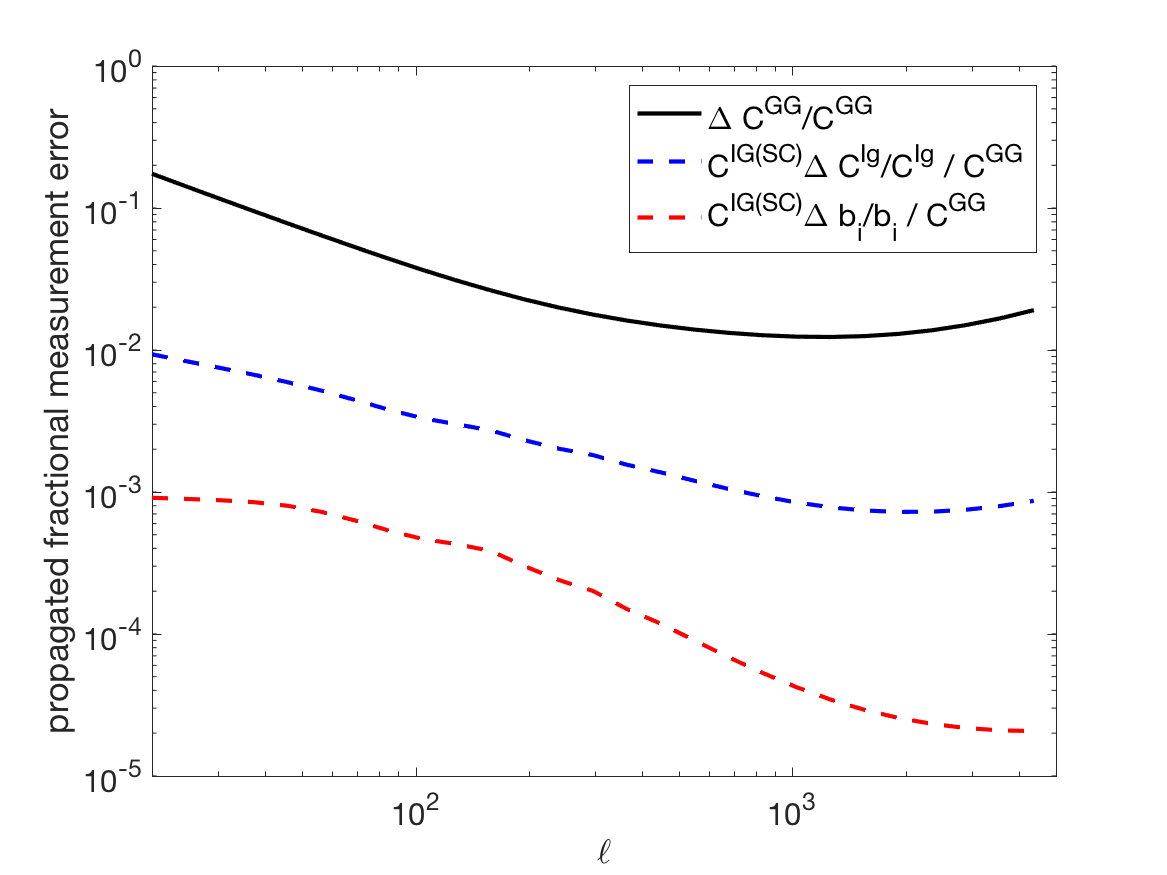}
\caption{Statistical error of SC. Solid black curve shows the statistical error of $C^{GG}$. Dashed blue curve shows the extra statistical error introduced by measuring $C^{Ig}$ using SC. Dashed red shows the extra statistical error introduced by measuring galaxy bias $b_i$
\label{statistical error}}
\end{figure}

One needs to keep in mind that, after applying SC, there are two different kinds of uncertainties. First, the systematic/residual bias, caused by the small residual part that the approximation of Eq.~(\ref{scaling-1}) fails to clean in the IG contamination, and also by the small GI/II contamination from the particular binning method of SC. Second, the statistical error [introduced by propagating the error from extra measurements Eq.~\eqref{scaling-2}]. Two examples of residual bias are shown in the top-left panel of Fig~\ref{Reproduce Zhang}, where
\begin{equation}
\epsilon_{ij}\equiv  \frac{C^{IG}_{ij}-C^{IG}_{ij}|_{SC}}{C^{IG}_{ij}|_{SC}} \label{epsilon}
\end{equation}
is the fractional residual bias. We can see that SC works very well and can clean most of the $C^{IG}$ contaminations on far-away bins (e.g. i=1 j=7). But the residual error is larger on adjacent bins (e.g. i=3, j=4) with fractional residual bias typically $\lesssim10\%$. \textcolor{black}{In the future for more advanced surveys, tomographic bin-size smaller than $\Delta_z^P=0.2$ may be required. We have also tested the value of $\epsilon_{ij}$ in that case, and it remains $<0.1$ and even smaller than their counterpart with a larger bin-size.} The extra statistical error is shown in Fig.~\ref{statistical error}, with the propagated fractional measurement error $f^{thresh}_{ij}\equiv \Delta C^{IG}_{ij}/C^{GG}_{ij}$ shown in blue/red for different sources of noise. Blue comes from the $C^{Ig}$ measurement introduced in E.\,\eqref{scaling-2}, and red comes from the $b_i$ measurement. The expressions for these two error are given in Ref.\,\cite{SC.Zhang} and Eq.\,\eqref{shot error 1} and \eqref{shot error 2} in our Appendix. The fractional measurement error without applying SC is $e^{min}_{ij}\equiv \Delta C^{GG}_{ij}/C^{GG}_{ij}$ and is shown in black.

\textcolor{black}{In this work we are not including the detailed analysis of the uncertainty in $Q_i$ measurement. This uncertainty fully comes from the uncertainty in photo-z, based on its definition in Eq.\,\eqref{Q}. According to the photo-z prior we choose (see Subsection \ref{subsection photo-z}), we perturb the photo-z parameters with their 1-$\sigma$ prior (i.e. $\delta\Delta_z=0.005$ and $\delta\sigma_z=0.006$ ), and estimated the uncertainty in $Q_i$ to be at $\sim0.1Q_i$ level. By propagating this error through Eq.\,\eqref{scaling-2} and \eqref{scaling-1}, the extra measurement uncertainty $\Delta C^{IG}$ is estimated to the order of $\sim0.1C^{IG}$, which is about two orders of magnitude smaller than $\Delta C^{GG}$. Thus we safely ignore this uncertainty in this analysis and leave the detailed exploration of this topic for future studies.}

We have seen examples in Fig.~\ref{Reproduce Zhang} that after using SC, the residual bias is expected to be at $\epsilon_{ij}<0.1$ level. Fig.~\ref{statistical error} shows that the extra statistical error introduced by using SC is at $\sim 0.1\Delta C^{GG}_{ij}$ level. Thus by using SC we clean most of the IG signal while introducing a reasonably small error. Detailed discussion for these two uncertainties and how they will affect the forecast of cosmological parameters will be included in the next section and the Appendix.

By calculating $Q_i(l)$, $C^{Ig}$ is extracted from $C^{(2)}$ via Eq.\,\eqref{scaling-2} (which is based on the the difference between Ig and Gg signals). In this step we use the information in photo-z to separate the IA part from the shear part. So extracting the GI signal is reduced to applying Eq.~(\ref{scaling-1}) and Eq.~(\ref{scaling-2}), with the perfect IA model already embedded within the measured $C^{Ig}$ spectra. The efficiency of SC depends on the quality of $W_{ij}\Delta_i/b_i$ in Eq.~(\ref{scaling-1}) and the quality of $Q_i$ for photo-z, but it is independent of the underlying IA model. This feature differs from the marginalization method of IA, whose efficiency does depend on the IA model.

We do not use the auto-spectra ($i=j$), in order to avoid the fact that in auto-spectra SC has relatively lower accuracy for the IG signal, plus the effect of the II signal is non-negligible. Because of this, some constraining power is lost. We showed that SC still does well only with cross-spectra ($i\ne j$).

In this work we will show that SC is competitive with the marginalization method of modeling IA. Furthermore, there is also potential of complement between SC and modeling IA. Modeling IA relies on deep understanding of the physics of IA, which is expected to be improved in the next decade. SC can model-independently measure IG spectra with a quite good accuracy when the two correlated bins are far away, for example $i=1, j=7$; see the top-left panel of Fig.~\ref{Reproduce Zhang}. The measurement of IA using SC can be used as a guide for IA modeling.

In this work, all the SC terms are calculated using our own FORTRAN/Python code. All the power spectra and models for IA, photo-z and galaxy bias are generated using CosmoSIS\cite{CosmoSIS}. The Fisher formalism is calculated using our own Matlab code. Some details about the internal infrastructure are introduced in the appendix.

\section{Forecasting the SC Performance (Fisher Formalism)} \label{Section Fisher}

A Fisher matrix is built to estimate the constraints on the cosmological parameters by propagating the uncertainties of measurements to the uncertainties of the cosmological parameters. In this work we will compare the results of the forecast for LSST/WFIRST/Euclid under 3 cases:

(A) a pure cosmic shear case with no systematics in the signal, assuming no IA, with the observed signal $C_{ij}=C^{GG}_{ij}+\delta_{ij}C^{GG,N}_{ij}$. This no-systematics case is the ideal scenario to use as a reference case for comparisons.

(B) Used for SC. Use the SC to subtract IA signal, $C_{ij}=C^{(1)}_{ij}-C^{IG}_{ij}|_{SC}$, also marginalize over the photo-z parameters.

(C) IA signal added by modeling will serve for the marginalization method. Marginalize over the IA parameters, $C_{ij}=\hat{C}^{(1)}_{ij}$, and marginalize over the photo-z parameters.

The above three scenarios will be discussed in detail in the subsections that follow.

\subsection{Reference Ideal Case (No Systematics)}

In this case, we show the Fisher Formalism for the ideal scenario with no presence of any kind of systematics. For one redshift bin (no tomography) and a given signal $C$, the Fisher matrix is expressed as \cite{Ishak2005}
\begin{equation}
F_{\alpha\beta}=\sum\limits_{\ell}\frac{f_{\rm sky}(2\ell+1)}{2}\left(\frac{1}{C^{GG}+C^{GG,N}}\right)^2\frac{\partial C^{GG}}{\partial p^\alpha}\frac{\partial C^{GG}}{\partial p^\beta}~,
\end{equation}
in which $p^\alpha,p^\beta\in$ \{ wCDM parameters + nuisance parameters \}. $C^{GG}$ is the shear power spectrum. $C^{GG,N}$ is the random shape noise spectrum, defined as $C^{GG,N}=4\pi f_{\rm sky}\gamma_{\rm rms}^2/N_i$, with the rms shape error $\gamma_{\rm rms}=0.26$ for LSST/WFIRST/Euclid and the number of galaxies in the i-th redshift bin $N_i$, while in this non-tomography case we only have one redshift bin. Considering tomography, the Fisher matrix is \cite{Hu1999}, with our notation:
\begin{widetext}
\begin{equation} \label{Fisher Tomo}
F_{\alpha\beta}=\sum\limits_{\ell}\frac{f_{\rm sky}(2\ell+1)}{2}Tr[(C_{\rm model}+C^{GG,N})^{-1}\frac{\partial C_{\rm model}}{\partial p^\alpha}(C_{\rm model}+C^{GG,N})^{-1}\frac{\partial C_{\rm model}}{\partial p^\beta} ]~,
\end{equation}
which comes from the likelihood $\mathcal{L}=e^{-\frac{1}{2}\chi^2}$, with $\chi^2$ defined as:
\begin{equation}
\chi^2=\sum\limits_{\ell}\frac{f_{\rm sky}(2\ell+1)}{2}Tr[(C_{\rm model}+C^{GG,N})^{-1}(C_{\rm model}-C_{\rm data})(C_{\rm model}+C^{GG,N})^{-1}(C_{\rm model}-C_{\rm data})]~.
\end{equation}
\end{widetext}

Here all the $C$'s are $n_{\rm bin}\times n_{\rm bin}$ matrices, where $n_{\rm bin}$ is the number of tomographic bins.

For cosmic shear, when no intrinsic shape is assumed, $C_{\rm model}=C^{GG}$. When binning in $\ell$ is applied, the quantity $2\ell+1$ needs to be substituted by $(2\ell+1)\Delta\ell \sim 2\ell\Delta\ell$, for LSST $\Delta\ell=0.2\ell$.

\subsection{Using Self-Calibration} \label{subsection using SC}

The SC case is the one we are going to focus on in this work. Here Self-Calibration is used to subtract the $C^{IG}_{ij}$ signal while minimizing $C^{GI}_{ij}$ and $C^{II}_{ij}$, as introduced in Section~\ref{Section SC}. So the measured GG spectrum with SC is
\begin{equation}\label{GG(SC)}
C_{\rm data}=C^{GG(SC)}_{ij}=C^{(1)}_{ij}-C^{IG(SC)}_{ij}-C^{GG,N}_{ij}~,
\end{equation}
where $C^{GG(SC)}_{ij}$ denotes the GG spectrum measured by SC, with both the SC-measured IG spectrum $C^{IG(SC)}_{ij}$ and the noise spectrum from the observed shear spectrum subtracted.

If we assume the binning method of SC is constructed so that $C^{GI}_{ij}$ and $C^{II}_{ij}$ vanish, and considering that SC cleans most of the $C^{IG}_{ij}$ signal, as shown in top-left panel of Fig.~\ref{Reproduce Zhang}, then the theoretical form for Eq.~\eqref{GG(SC)} becomes:
\begin{equation} \label{approxGG}
C_{\rm model}=C^{GG}_{ij}+C^{IG}_{ij}-C^{IG(SC)}_{ij}\approx C^{GG}_{ij}.
\end{equation}

We note that the quality of Eq.~\ref{approxGG} depends on the efficiency of the SC, which includes the accuracy of the $C^{IG}_{ij}$ measurement in Eq.~\ref{scaling-1} as well as the effect of $C^{GI}_{ij}$ and $C^{II}_{ij}$ after binning. In Ref.~\cite{SC.Zhang}, it has been shown that SC only applies to cross-spectra ($i\ne j$) because of the failure of the binning method in dealing with $C^{II}_{ii}$. We also obtained a lower accuracy of $C^{IG(SC)}_{ii}\sim 0.8C^{IG}_{ii}$ in the auto-spectra. Therefore SC does not apply in the auto-spectra, which means as a price of using SC, some constraining power is lost, but much smaller than the nulling technique methods \cite{Joachimi2008,Joachimi2009,Joachimi2010},. In our results we will additionally show that the effect of $C^{II}_{ij}$ in adjacent bins ($j=i+1$) will also cause some small bias.

For SC we use another form of Eq.~(\ref{Fisher Tomo}), that is:
\begin{align}\label{Fisher Cov}
F_{\alpha\beta}=f_{\rm sky}\sum\limits_{l}\sum\limits_{ijpq}C_{ij,\alpha}[Cov(C,C)]^{-1}_{(ij),(pq)}C_{pq,\beta},
\end{align}
where
\begin{equation}
Cov(C_{ab},C_{cd})=\frac{1}{{2l\Delta l f_{\rm sky}}}(C_{ac}C_{bd}+C_{ad}C_{bc})
\end{equation}
is the covariance of the spectra. This expression is also given in Hu \& Jain 2004 \cite{Hu2004} and Clerkin et al \cite{Clerkin2014}.

Firstly, for SC, the terms in Eq.~(\ref{Fisher Cov}) need to be carefully discussed. For the forecast work, we calculate $\hat{C}^{(1)}$ from $\hat{C}^{(1)}_{ij}={C}^{GG}_{ij}+{C}^{IG}_{ij}+{C}^{GI}_{ij}+{C}^{II}_{ij}+\delta_{ij}C^{GG,N}_{ii}$, as shown in Eq.~\eqref{observables}, with the IA spectra resulting from a given IA model. Recall that SC is a technique that measures the IG spectrum without assuming a model of the IA, the partial derivative indexes, $\alpha$ and $\beta$, should not depend on the IA model. Thus according to Eq.~\eqref{approxGG}, we assume $C_{ij,\alpha}=C^{GG}_{ij,\alpha}$ so that SC removes all the IA signal. The residual IA spectra after applying SC is a small fraction, and will result in a shift in the best-fit cosmology that we will discuss later.

Secondly, we need to derive the covariance of the signal, which is based on the deviation of the measured $C^{GG(SC)}$ from the theoretical $C^{GG}$, which we give as:
\begin{widetext}
\begin{align} \label{SC Cov}
&Cov(C^{GG(SC)}_{ij},C^{GG(SC)}_{pq})=Cov(C^{(1)}_{ij}-C^{IG(SC)}_{ij},C^{(1)}_{pq}-C^{IG(SC)}_{pq}) \\\notag
&=Cov(C^{(1)}_{ij},C^{(1)}_{pq})-Cov(C^{(1)}_{ij},{C}^{IG(SC)}_{pq})-Cov({C}^{IG(SC)}_{ij},C^{(1)}_{pq})+Cov(C^{IG(SC)}_{ij},C^{IG(SC)}_{pq}).
\end{align}
\end{widetext}

\textcolor{black}{In Eq~\eqref{SC Cov}, the first term is well known as the covariance of the observed cosmic shear. The other three terms contain the impact of the extra statistical error (as shown in Fig.\,\ref{statistical error}) introduced by SC in the previous section. The 4th term gives the pure influence of the extra measurement errors, while the 2nd and 3rd terms give the correlation between the error of the observed cosmic shear and the extra measurement errors. The uncertainty in photo-z are not included in this covariance. They will be addressed during the marginalization of the photo-z parameters.} The effect of the residual bias will also cause the shift of the best-fit cosmological parameters. This will be discussed in Subsection~\ref{Section shift}. The overall contribution of 2nd, 3rd and 4th terms of the Covariance to the Fisher matrix are negligible compared to the dominating 1st term. This agrees with Zhang's \cite{SC.Zhang} argument that the error introduced by SC is much smaller than that of the cosmic shear. The detailed derivation of Eq.\,\eqref{SC Cov} is included in the Appendix.

The final expression of the Fisher matrix for SC is then

\begin{align}\label{Fisher SC}
F_{\alpha\beta}=\sum\limits_{l}\sum\limits_{ijpq}C^{GG}_{ij,\alpha}[Cov(C^{GG(SC)},C^{GG(SC)})]^{-1}_{(ij),(pq)}C^{GG}_{pq,\beta}.
\end{align}

\subsection{Marginalization over IA and photo-z parameters} \label{subsection marginalization}

This scenario is similar to ``No Systematics''. But with the existence of the IA, the model we use needs to take the IA spectra into account: $C_{\rm model}=C^{GG}+C^{IG}_{\rm model}+C^{GI}_{\rm model}+C^{II}_{\rm model}$.

The observable $C_{\rm data}=C^{(1)}$ is given by Eq.~(\ref{observables}). The Fisher matrix with tomography is in the same form as Eq.~(\ref{Fisher Tomo}). For this marginalization over IA parameters method, we need to include the IA parameters such that $p^\alpha,p^\beta\in$ \{ wCDM parameters + photo-z parameters + IA parameters\}. The IA model is discussed in the next section.

\textcolor{black}{Note that the extra observables introduced in Eq.\,\eqref{C2} and \eqref{C3} are for the SC case, only in order to get the IA signal measurement and galaxy bias measurement, rather than being used to get better constraints on cosmological parameters. Neither the SC case nor the marginalization case uses the constraining power from galaxy-galaxy lensing or galaxy-galaxy clustering in th present analysis. Also we want to emphasize that in this marginalization case, all the spectra have been used, including the ($i=j$) auto-spectra, which are not being used due to the limitation of SC we discussed in the previous section.}

\section{Photo-z, galaxy bias and IA models} \label{Section models}

In this section we introduce the models of photo-z, galaxy bias and IA being used in our work. The SC technique itself doesn't assume any model of the IA, but this is needed for the marginalization method. The information for IA ($C^{IG(SC)}$) completely comes from $C^{Ig}$, which is obtained by Eq.~(\ref{scaling-2}), utilizing the information in photo-z.

\subsection{Photo-z Model} \label{subsection photo-z}

Photo-z is another major problem in photometric surveys \cite{Ma2006photoz}, therefore we must take its effect into consideration. The overall true-z distribution of the survey and photo-z (Gaussian) probability distribution function are expressed as:
\begin{align} \label{redshift}
n(z)&\propto z^\alpha {\rm exp}\left[-(\frac{z}{z_0})^\beta\right],\\
p(z^P|z)&=\frac{1}{\sqrt{2\pi\sigma_z(1+z)}}{\rm exp}\left[-\frac{(z-z^P-\Delta_z^i)^2}{2(\sigma_z(1+z))^2}\right].
\end{align}

Here $z$ is the true redshift, $z^P$ is the photo-z. The normalized redshift distribution for each tomographic bin is expressed as $n_i(z)$, which is given by:
\begin{align} \label{n_i}
n_i(z)=\frac{\int_{z_{i,\rm min}^P}^{z_{i,\rm max}^P} n(z)p(z^P|z)dz^P}{\int_0^\infty[\int_{z_{i,\rm min}^P}^{z_{i,\rm max}^P} n(z)p(z^P|z)dz^P]dz}.
\end{align}

For different surveys, different specifications are applied. We choose photo-z scatter $\sigma_z=0.05$, redshift bias $\Delta_z=0$ as common specifications, with assumed Gaussian priors: ${\rm Gaussian}(0, 0.005)$ (a Gaussian distribution with mean value 0 and 1-$\sigma$ uncertainty 0.005) for $\Delta_z$ and ${\rm Gaussian}(0.05, 0.006)$ for $\sigma_z$ \cite{Krause2016}. The other specifications for LSST, Euclid and WFIRST are shown in Table\,\ref{surveys} \cite{10bins,Euclid}.

\begin{table}[h]
	\caption{Survey Parameters. We estimate in what follows the performance of the IA self-calibration method for the 3 cosmic shear surveys.}\label{surveys}
	\begin{ruledtabular}
		\begin{tabular}{ c c c c c c c c }
			 & $f_{\rm sky}$ & $\gamma_{\rm rms}$ & $n_{\rm eff}$ & $z_0$ & $\alpha$ & $\beta$ & $z_{\rm max}$ \\
			 \hline
			LSST & 0.436 & 0.26 & 26 & 0.5 & 1.27 & 1.02 & 3.5 \\
			WFIRST & 0.053 & 0.26 & 45 & 0.6 & 1.27 & 1.02 & 4.0 \\
			Euclid & 0.364 & 0.26 & 30 & 0.6374 & 2 & 1.5 & 2.5 \\
		\end{tabular}
	\end{ruledtabular}
\end{table}

\begin{figure}[t!]
\includegraphics[width=0.9\columnwidth]{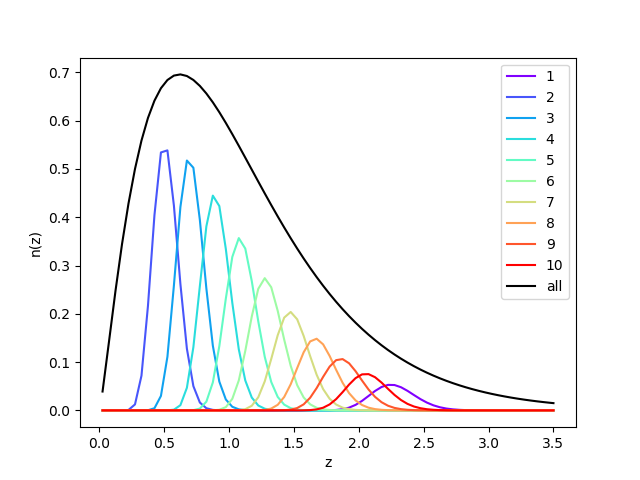}
\caption{Redshift distribution with 10 tomographic bins of SC. Photo-z range $[0.4,2.4)$ with bin-width $\Delta z^P=0.2$. This is the major part of Weak Lensing constraining power. In this plot the redshift distributions are not normalized, so that they are calculated by the numerator of Eq.~\eqref{n_i}. This is for convenient comparison with the overall distribution.
\label{fig n(z)}}
\end{figure}

The binning method also needs to be carefully considered. Bridle \& King (2007)\cite{BridleKing} showed that more bins are required for cosmological parameter estimation when considering IA. In Heymans et al (2013) \cite{CFHTLenSHeymans}, the instability of covariance matrix with a large bin-number $n_{\rm bin}$ has also been discussed. To strike a balance between these two aspects, we choose $n_{\rm bin}=10$. This choice of $n_{\rm bin}$ agrees with the binning in Schaan et al (2016) \cite{10bins} for LSST, Euclid and WFIRST. The SC technique requires appropriate choice of photo-z bin-width, hence we choose the bin-width $\Delta z^P=0.2$ consistent with Zhang \cite{SC.Zhang}. Fig.~\ref{fig n(z)} shows the unnormalized redshift distribution of 10 tomographic bins for LSST with photo-z range $0.4<z^P<2.4$. For Euclid and WFIRST we choose the same binning: $0.4<z^P<2.4$ with bin-width $\Delta z^P=0.2$.

\subsection{Galaxy Bias Model}

The SC technique measures the averaged galaxy bias using
\begin{equation}
C^{gg}_{ii}(l)\approx b_i^2(l)C^{mm}_{ii}(l) \label{data measured b_i}
\end{equation}
for real data, where the matter angular power spectrum $C^{mm}$ is expected to come from CMB experiments as it's tightly constrained according to Ref.\,\cite{SC.Zhang}. However the tension between cosmic shear and CMB \cite{CFHTLenSHeymans,Planck2016,Hildebrandt2016,Leauthaud2017,Lin2017} will need to be taken into account . In this forecast work we assess the bias by averaging over each redshift bin:
\begin{equation}
b_i=\int_0^\infty b_g n_i dz , \label{b_i}
\end{equation}
which requires a bias model of $b_g$. \textcolor{black}{We numerically tested Eq.\,\eqref{data measured b_i} and \eqref{b_i} and found them to be almost the same.}

We use the Generalized Time Dependent (GTD) Bias Model \cite{Clerkin2014}, which has been shown in linear scale as an encapsulation of several time-dependent models and has good agreement with simulations \cite{Clerkin2014}. The expression for the GTD model is given by:
\begin{align}
b(z)=c+(b_0-c)/D^\alpha(z) \label{GTD model},
\end{align}
in which $D(z)$ is the linear growth function, satisfying $\ddot{D}+2H(z)\dot{D}-\frac{3}{2}\Omega_m H_0^2(1+z)^3D=0$. The parameters we use are $c=0.57$, $b_0=0.79$ and $\alpha=2.23$ \cite{Clerkin2014}.

\textcolor{black}{The scaling relation Eq.~(\ref{scaling-1}) of SC is not sensitive to the choice of bias model, as the galaxy bias enters both $C^{Ig}$ and $b_i$ and hence roughly cancel. The uncertainty being introduced (shown in Eq.\,\eqref{shot error 2}) by measuring the weighted mean galaxy bias $b_i$ is also easier to constrain compared to marginalizing over the bias parameters in Eq.\,\eqref{GTD model}. We showed in Fig.\,\ref{statistical error} that this part of extra measurement uncertainty is negligible compared with the other two kinds. Note here the bias model is used to generate the galaxy bias signal, while the observable $b_i$ rather than the galaxy bias $b_g$ is used in this analysis for SC. Thus there is no need to marginalize over the bias parameters in either the SC case or the IA marginalization case.}

\subsection{IA model} \label{subsection IA model}

The expression for components in Eq.~(\ref{observables}) are given by Ref.\,\cite{BridleKing,Hirata2003}
\begin{subequations}
\begin{align}
C^{GG}_{ij}&=\int_0^\infty\frac{q_i(\chi)q_j(\chi)}{\chi^2}P_\delta(k;\chi)d\chi, \\ \label{GG}
C^{IG}_{ij}&=\int_0^\infty\frac{n_i(\chi)q_j(\chi)}{\chi^2}P_{\delta,\gamma^I}(k;\chi)d\chi, \\ \label{IG}
C^{GI}_{ij}&=\int_0^\infty\frac{q_i(\chi)n_j(\chi)}{\chi^2}P_{\delta,\gamma^I}(k;\chi)d\chi, \\ \label{GI}
C^{II}_{ij}&=\int_0^\infty\frac{n_i(\chi)n_j(\chi)}{\chi^2}P_{\gamma^I}(k;\chi)d\chi. \\ \label{II} \notag
\end{align}
\end{subequations}

\begin{table}[h]
	\caption{Fiducial Cosmological Model}\label{FiducialModel}
	\begin{ruledtabular}
		\begin{tabular}{ c c c c c c c }
			$\Omega_m$ & $h_0$ & $\sigma_8$ & $n_s$ & $\Omega_b$ & $w_0$ & $w_a$ \\
			0.315  &  0.673  &  0.829  &  0.9603  & 0.049 &  -1.0  &  0 \\
		\end{tabular}
	\end{ruledtabular}
\end{table}

\begin{figure*}[t!]
	\includegraphics[width=2.2\columnwidth]{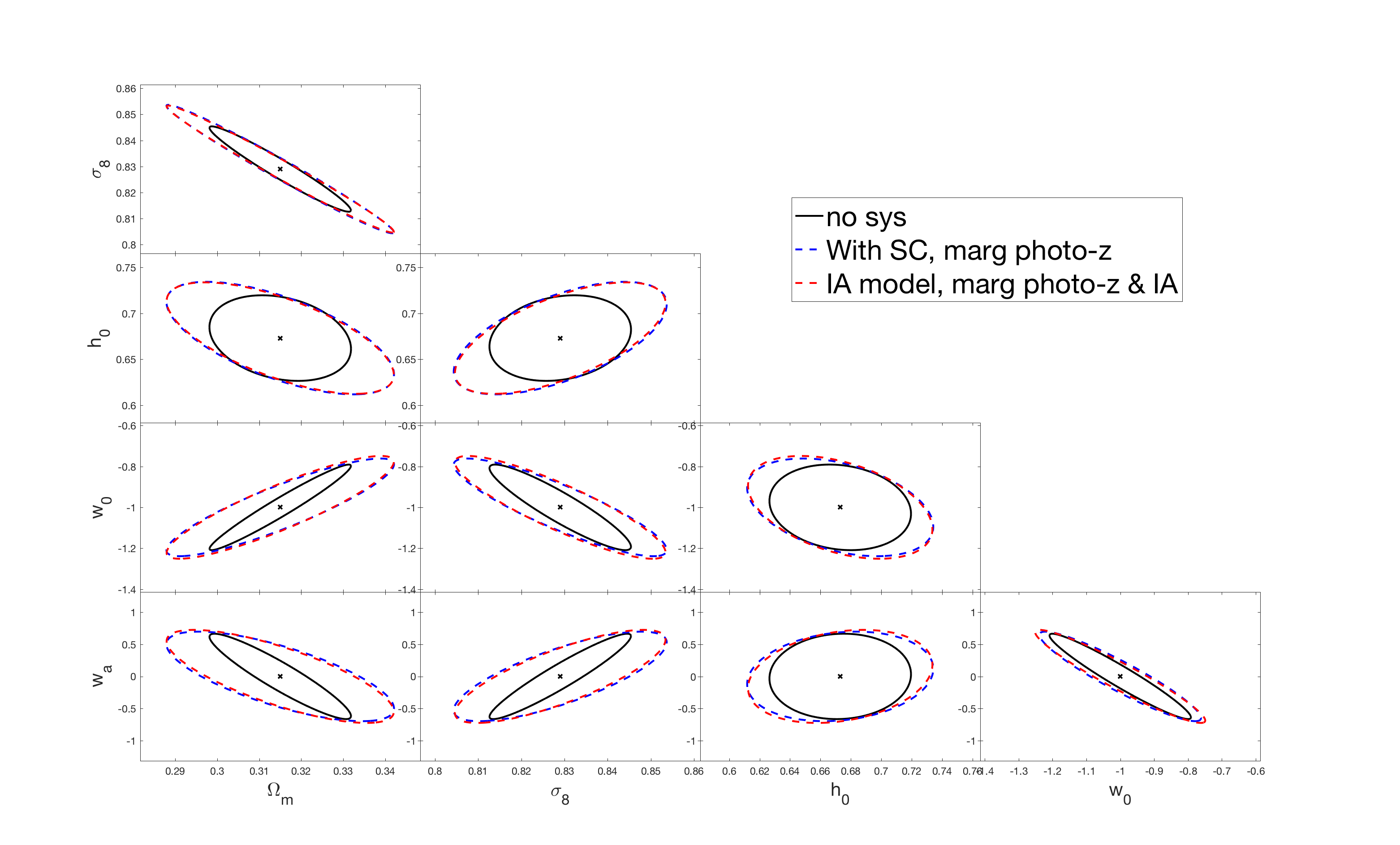}
	\caption{Additional errors from applying IA mitigation methods and marginalization over the photo-z's: $68\%$ Confidence Contours of Cosmological Parameters with a fixed fiducial cosmology. The black contours are the ideal cases for LSST where no systematics are assumed. The blue dashed contours are using SC to mitigate IA signal with marginalization over the photo-z parameters. The red dashed contours use the marginalization method for both IA and photo-z, with the models introduced in text. The central cross-dots are the assumed fiducial cosmology as shown in Table \ref{FiducialModel}.
		\label{contour}}
\end{figure*}

$P_\delta(k;\chi)$ in Eq.~(\ref{GG}) is the (non-linear) matter power spectrum at the redshift of the lens. $n_i(\chi)$ is the redshift distribution in the i-th redshift bin. $q_i(\chi)$ is the lensing efficiency function for lens at $\chi_L(z_L)$ for the i-th redshift bin, written as
\begin{equation}
q_i(\chi_L)=\frac{3}{2}\Omega_m\frac{H_0^2}{c^2}(1+z_L) \int_{\chi_L}^\infty n_i(\chi_S)\frac{(\chi_S-\chi_L)\chi_L}{\chi_S}d\chi_S.
\end{equation}

The two relevant 3-D IA spectra are
\begin{subequations}
\begin{align} 
P_{\delta,\gamma^I}=-A(L,z)\frac{C_1\rho_{m,0}}{D(z)}P_\delta(k;\chi), \label{IA 3D 1} \\
P_{\gamma^I}=A^2(L,z)(\frac{C_1\rho_{m,0}}{D(z)})^2P_\delta(k;\chi). \label{IA 3D 2}
\end{align}
\end{subequations}

In Eq.\,\eqref{IA 3D 1} and \eqref{IA 3D 2}, $\rho_{m,0}=\rho_{crit}\Omega_{m,0}$ is the mean matter density of the universe at $z=0$. $C_1=5\times 10^{-14}(h^2M_{\rm sun}/Mpc^{-3})$ is derived by comparing with SuperCOSMOS \cite{BridleKing}. We use $C_1\rho_{crit}\approx 0.0134$ as in Ref.~\cite{Krause2016}.

A commonly used IA model is the redshift- and luminosity- dependent IA amplitude model $A(L,z)$:
\begin{equation}
A(L,z)=A_0(\frac{L}{L_0})^\beta(\frac{1+z}{1+z_0})^\eta.
\end{equation}

The IA parameters for this model are $A_0$, $\beta$ and $\eta$, with $z_0=0.3$ the (observationally motivated) pivot redshift, and $L_0$ the pivot luminosity corresponding to an absolute magnitude of -22 in r-band.

In this work, we fix $\beta=0$ and leave the dependence on luminosity for future work. We don't require a significantly complicated model as it has been shown that in CFHTLenS data \cite{Joudaki2016} and KiDS-450 data \cite{Hildebrandt2016}, the IA signal is insensitive to any luminosity-dependence for their surveys. In future work we will explore the effect of a varying $\beta$, as well as the impacts of different luminosity functions of red- and blue-type galaxies. The amplitude of IA is chosen to be $A_0=1$.

%

\section{Results} \label{Section results}

\begin{figure*}[t]
	\includegraphics[width=0.65\textwidth]{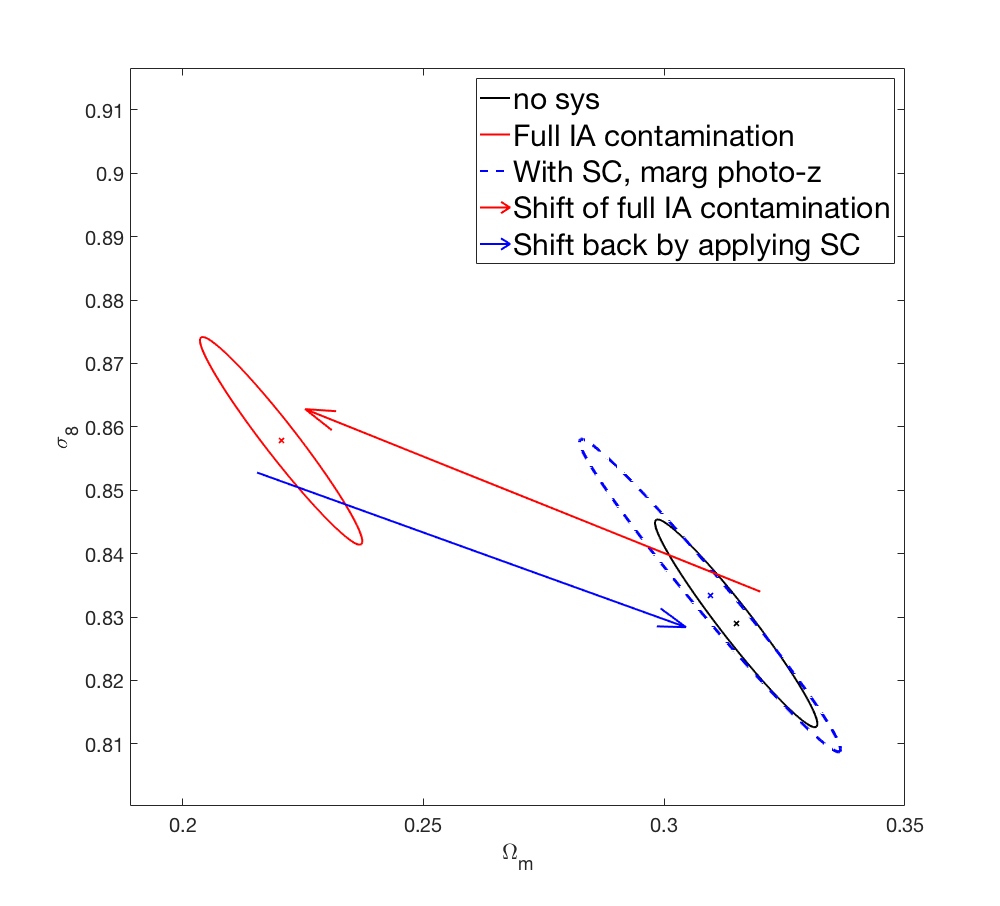}
	\caption{An example of the shift in cosmological parameters in $\Omega_m$ and $\sigma_8$ space. Black contour is the no systematics case. If we do not deal with IA at all, the IA contamination will lead to a shift in the cosmological parameters, shown as the red contour with the red arrow. The shift of red contour is calculated by the Newtonian method we introduced in Eq.\,\eqref{Newtonian shift}, but with the full IA contamination (not dealing with IA). This shift agrees with the shift using different IA models in Ref.\,\cite{Krause2016}. When applying SC to clean the IA signal, and marginalize over the photo-z parameters, the best-fit cosmological parameters will be brought back to the blue dashed contour (data from Fig.\,\ref{shift with II} which will be shown later) along the blue arrow.
		\label{shift_example}}
\end{figure*}

\begin{figure*}[t]
	\includegraphics[width=2.0\columnwidth]{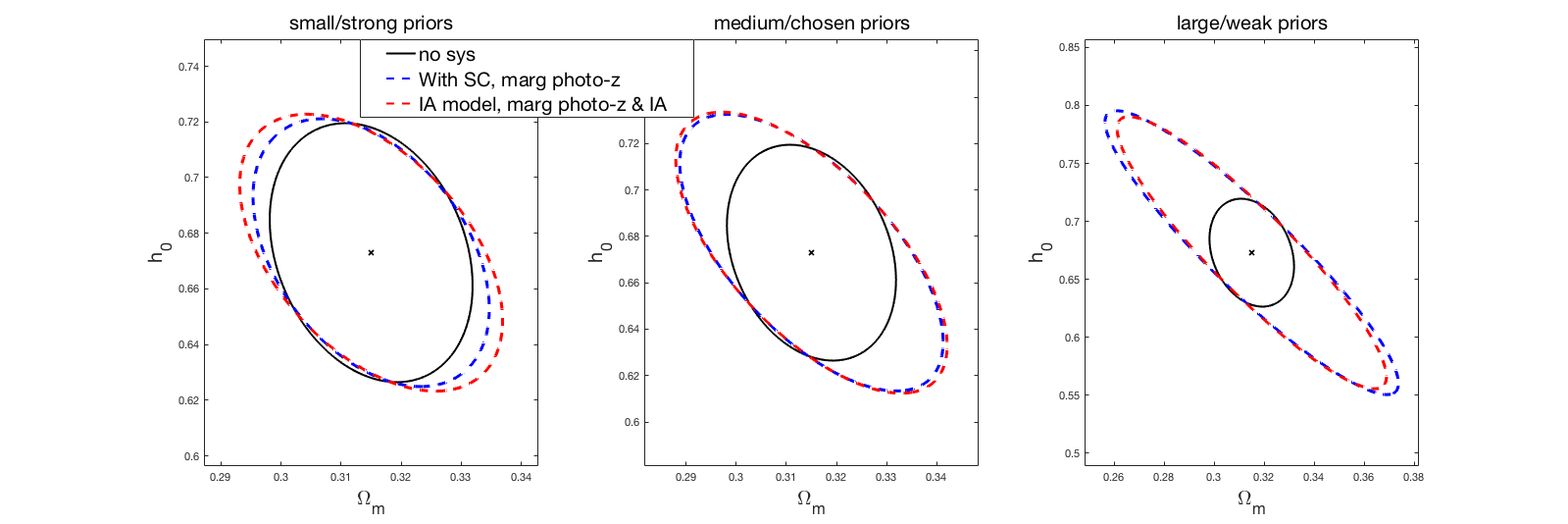}
	\caption{\textcolor{black}{An illustration of how SC is moderately more sensitive to the choice of photo-z priors than the marginalization method. The middle panel is the same of the $h_0-\Omega_m$ plot in Fig.\,\ref{contour} (but re-scaled here for display purposes), with the photo-z priors of  ${\rm Gaussian}(0, 0.005)$ in $\Delta_z$ and ${\rm Gaussian}(0.05, 0.006)$ in $\sigma_z$ (see text for notation). The left panel shows the contours with stronger photo-z priors of ${\rm Gaussian}(0, 0.002)$ in $\Delta_z$ and ${\rm Gaussian}(0.05, 0.003)$ in $\sigma_z$. The right panel shows the contours with weaker photo-z priors ${\rm Gaussian}(0, 0.1)$ in $\Delta_z$ and ${\rm Gaussian}(0.05, 0.1)$ in $\sigma_z$. As the photo-z prior becomes weaker, the contours become bigger for both the SC and the marginalization cases but the contour size increases slightly faster with the photo-z priors in the SC case than in the marginalization case. This shows that statistical error in the SC case has a moderately stronger dependence on the photo-z quality than the marginalization case.}
		\label{prior_dependence}}
\end{figure*}

\begin{figure*}[t]
	\includegraphics[width=2.2\columnwidth]{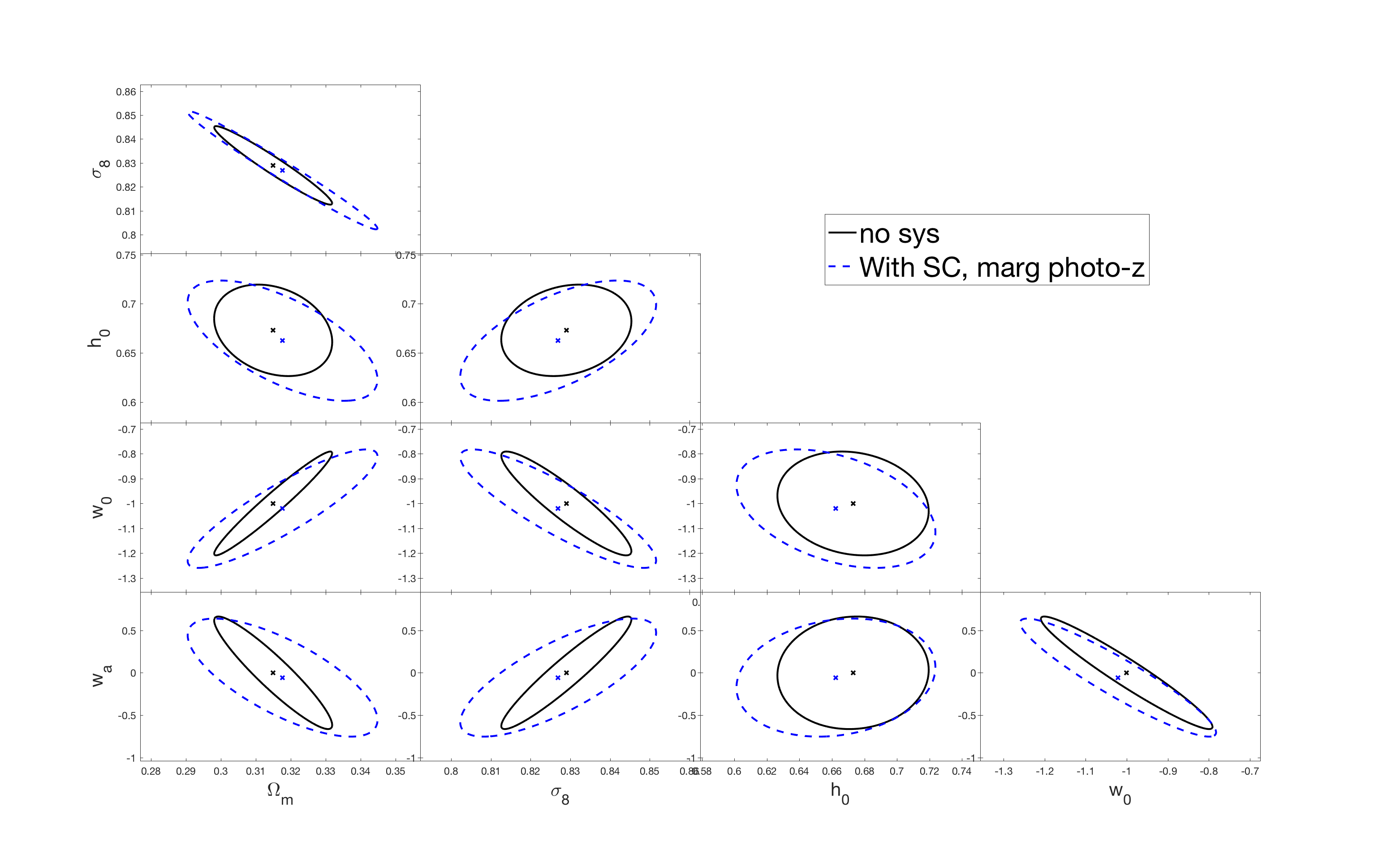}
	\caption{Residual shift of best-fit cosmology for LSST when assuming GI IA only (i.e. no II), and only using the cross-spectra to constrain  cosmological parameters, so that the shift in cosmology will be fully caused by the residual of GI in the cross-spectra. Solid black contours are the ideal no systematics situation, with the black central dots being the true fiducial cosmology. The best-fit cosmology using SC is shown in blue. All the residual shifts are within $1\sigma$ of the black uncertainties.
		\label{shift no II}}
\end{figure*}

\subsection{Confidence Contours at the fixed Fiducial Cosmology}

In this section we will show the results of the forecast of the cosmological parameters using the 3 different methods described in Section~\ref{Section Fisher}: No Systematics, Using SC, and Marginalizing over IA parameters. Compared to the case of No Systematics, although both methods will lose accuracy(larger confidence contours), it is expected that a smaller shift of the cosmological parameters will be achieved by solving the IA problem in two different ways compared to not dealing with IA at all. We first compare the precision of the three scenarios by using the same fiducial cosmological parameters in the Fisher matrix. In the next subsection, we will investigate the accuracy of using SC, i.e. the shift of the best fit cosmological parameters. We will compare our results to those of Ref.\,\cite{Krause2016} as they also looked into the precision and accuracy effects. The fiducial cosmological model we use is shown in Table~\ref{FiducialModel}.

Fig.~\ref{contour} shows the uncertainty contours for the cosmological parameters for LSST. The black contours are the results assuming IA has been completely taken care of, thus the GG signal is the pure cosmic shear signal. The blue contours show when most of the IA signals have been cleaned out using SC, and marginalizing over the photo-z parameters \{$\sigma_z,\Delta_z$\}. The red contours use the IA model with marginalization over the IA parameters \{$A_0,\eta$\} together with the photo-z parameters. Despite the fact that we are using a different method, our results in black and red curves are in good agreement with those in Ref.~\cite{Krause2016}. This is good news as sanity check and allows us to compare our new results reliably to this reference case and also to those of Ref.\,\cite{Krause2016} for the marginalization case.

Although the SC and marginalization are found to perform almost equally well for the given underlying IA signal, as shown in Fig.\,\ref{contour}, the SC method has various intrinsic advantages and constitute a competitive and supplementary alternative. One major advantage is that SC does not need to assume an IA model. We benefit from SC in that we can have a model-independent confidence that the majority of the IA signals are subtracted. \textcolor{black}{However, a disadvantage of SC is that even though it doesn't use the constraining power of other probes, it still requires some information to subtract the IA spectra. Thus, in the future a full comparison between SC and IA modeling in the context of a combined probes analysis will be important.}

The main sources of enlargement in contours comparing the SC case and the no systematics case include: the extra measurement errors that are included in 2nd, 3rd and 4th terms in Eq.\,\eqref{SC Cov}, the constraining power loss for not including auto-spectra ($i=j$) in the SC case, and the marginalization process over the photo-z parameters. The sources of enlargement in contours comparing the marginalization case and the no systematics case include: the marginalization process over the IA parameters and photo-z parameters.

\textcolor{black}{We want to further discuss the dependency on photo-z quality of both the SC case and the marginalization case. As mentioned earlier, the extra measurement error of SC is negligible, thus the enlargement in the contour size of SC is mainly due to photo-z marginalization. Since SC uses the information in photo-z through Eq.\,\eqref{Q} and \eqref{scaling-2}, it is reasonable to expect the performance of SC will be more sensitive to the quality of photo-z. For that we have tested different priors for photo-z to compare the contours of the SC case and marginalization case, see Fig.\,\ref{prior_dependence}. As illustrated in the figure, as the photo-z prior becomes weaker, the contours become bigger for both the SC and the marginalization cases but the contour size increases slightly faster with the photo-z priors in the SC case. This shows that statistical error in the SC case has a moderately stronger dependence on the photo-z quality than in the marginalization case. The choice of the photo-z prior presented in Fig.\,\ref{contour} coincidentally gives similar contours for the two different methods but that should not be the case otherwise.}

\subsection{Residual Bias of best-fit Cosmological Parameters} \label{Section shift}

\begin{figure*}[t]
\includegraphics[width=2.2\columnwidth]{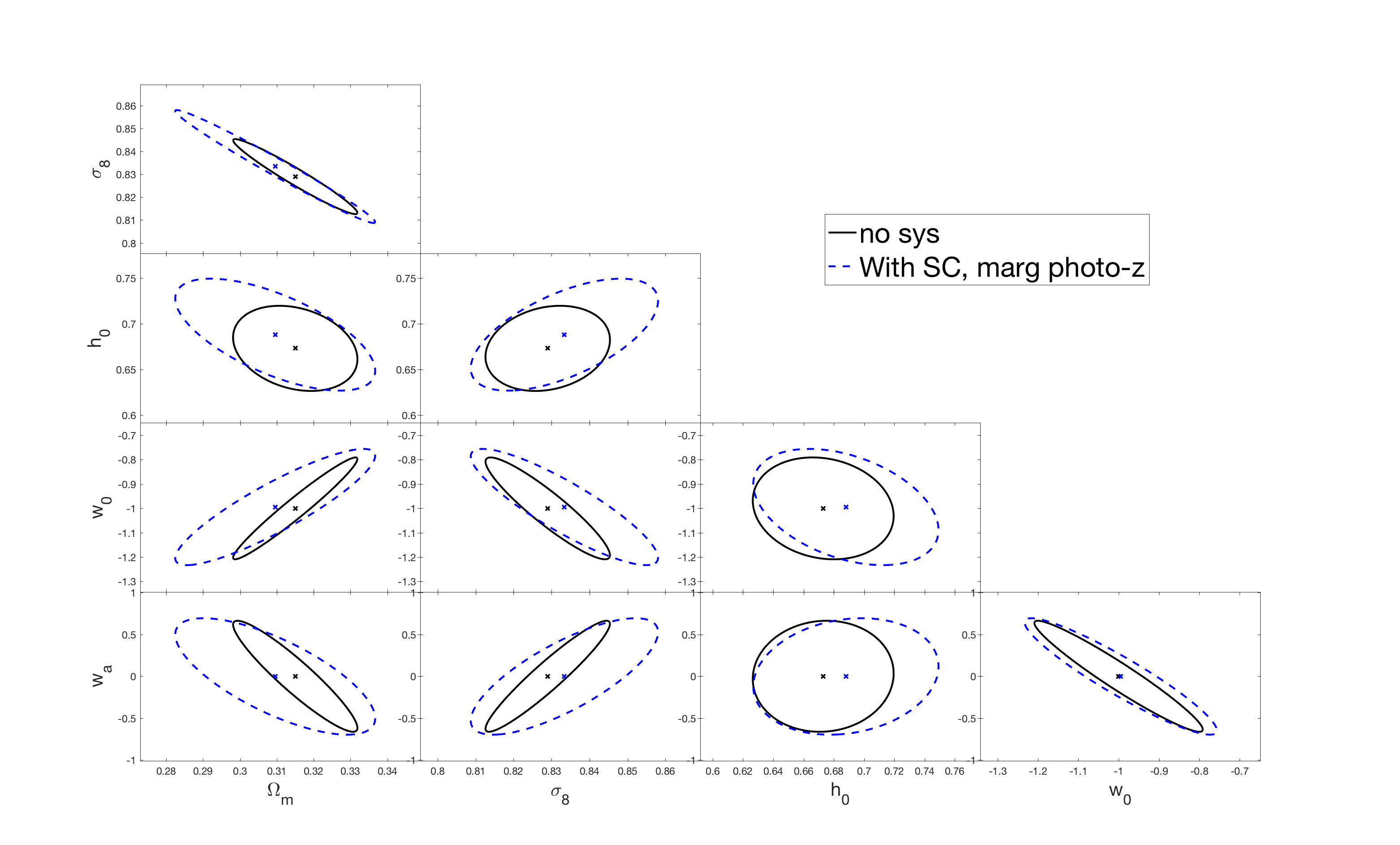}
\caption{Shift of the best-fit cosmology with the presence of II and GI in all spectra, also only using the cross-spectra to constrain the cosmological parameters for LSST. The shifts in cosmological parameters will be mainly caused by the residual GI and II in the cross-spectra. In non-adjacent bins, the residual GI dominates. In adjacent bins, II is more dominant than residual GI, causing a shift (blue) in the opposite direction compared to Fig.~\ref{shift no II}. All the shifts are within $1\sigma$ level of the black contours.
\label{shift with II}}
\end{figure*}

In this subsection we discuss the shift of the cosmological parameters due to the residual IA after applying SC. Fig.\,\ref{shift_example} gives an example of how the shift works. The existence of IA will result in a shift in the best-fit cosmological parameters, i.e. a biased estimation. By applying SC, this bias can be largely reduced as the major contamination is subtracted.

We employ a Newtonian method \cite{Kirk2012,Clerkin2014} to estimate the shift of the cosmological parameters from the true fiducial cosmology when using SC. We first assume a true fiducial Cosmological Model 1 (CM1), which is also the best-fit value for the 'no sys' scenario. The fiducial parameter values for CM1 are the ones we've been using in Table~\ref{FiducialModel}. We want to find a second Cosmological Model 2 (CM2) that maximizes the likelihood $e^{-\frac{1}{2}\chi^2}$ for SC, or minimizes $\frac{1}{2}\chi^2$ below.

\begin{widetext}
\begin{align}
\frac{1}{2}\chi^2&=\frac{1}{2}f_{\rm sky}\sum\limits_{l}\frac{(2l+1)\Delta l}{2}\left[C^{GG}-C^{GG(SC)}\right]_{(ij)}\left[Cov(C^{(1)}-C^{IG(SC)},C^{(1)}-C^{IG(SC)})\right]^{-1}_{(ij),(pq)}\left[C^{GG}-C^{GG(SC)}\right]_{(pq)}~,
\end{align}
in which $C^{GG(SC)}$ is the measured GG spectra using SC, defined in Eq.~\eqref{GG(SC)}.

By taking the partial derivative with respect to $p^\alpha$, we have:
\begin{equation}
\frac{1}{2}\frac{\partial \chi^2}{\partial p^\alpha}=f_{\rm sky}\sum\limits_{l}\frac{(2l+1)\Delta l}{2}\left[C^{GG}-C^{GG(SC)}\right]_{ij}\left[Cov(C^{(1)}-C^{IG(SC)},C^{(1)}-C^{IG(SC)})\right]^{-1}_{(ij),(pq)}\frac{\partial C^{GG}_{(pq)}}{\partial p^\alpha} \label{A_alpha}~.
\end{equation}
\end{widetext}

We set $\frac{1}{2}\frac{\partial \chi^2}{\partial p^\alpha}|_{\rm CM2}=0$ to solve for the best-fit values $p^\alpha|_{\rm CM2}$. Physically the shift $(P^\beta|_{\rm CM2}-P^\beta|_{\rm CM1})$ is caused by the residual IA spectra that is left after using SC. The shift should be small, so that we can expand as:
\begin{align}
0&=(\frac{1}{2}\chi^2)_{,\alpha}|_{\rm CM2}  \notag\\
&\approx (\frac{1}{2}\chi^2)_{,\alpha}|_{\rm CM1} + \sum\limits_{\beta}(\frac{1}{2}\chi^2)_{,\alpha\beta}|_{CM1}(P^\beta|_{\rm CM2}-P^\beta|_{\rm CM1}) \label{shift}.
\end{align}

Define $A_\alpha=(\frac{1}{2}\chi^2)_{,\alpha}|_{\rm CM1}$, and the shift $\Delta P_\beta=P_\beta|_{\rm CM2}-P_\beta|_{\rm CM1}$, while $(\frac{1}{2}\chi^2)_{,\alpha\beta}|_{\rm CM1}$ approximates the Fisher matrix $F_{\alpha\beta}$ of the SC method given in Eq.~(\ref{Fisher SC}). Hence we can solve for $\Delta P^\beta$ in the following matrix form:
\begin{equation} \label{Newtonian shift}
\Delta P^\beta\approx-(F^{-1})^{\alpha\beta}A_\alpha.
\end{equation}

\begin{table*}[t]
	\caption{\label{dp/dsigma} In this table we present the residual shift $\Delta P$ in cosmological parameters in terms of the 1-$\sigma$ uncertainties of the parameters as numerical details for Fig.\,\ref{shift no II} and \ref{shift with II}, to show the ideal and actual (with the contamination of II signals) performance of SC. The $\pm$ signs show the directions of the shifts. WE find that all the residual shits after applying the SC fall well-below the 1-$sigma$ level.}
	\begin{tabular}{  c | p{0.1\textwidth}<{\centering} | p{0.1\textwidth}<{\centering} | p{0.1\textwidth}<{\centering} | p{0.1\textwidth}<{\centering} | p{0.1\textwidth}<{\centering} | p{0.1\textwidth}<{\centering}  }
		\hline\hline
		\multirow{2}{*}{Survey} & \multirow{2}{*}{Method}
		& \multicolumn{5}{|c|}{$\Delta P/\sigma$} \\
		\cline{3-7}
		& & $\Omega_m$ & $h_0$ & $\sigma_8$ & $w_0$ & $w_a$ \\
		\hline
		\multirow{2}{*}{LSST}
		& with II & $-0.304$ & $0.372$ & $0.270$ & $0.035$ & $-0.005$ \\
		\cline{2-7}
		& without II & $0.150$ & $-0.261$ & $-0.126$ & $-0.133$ & $-0.124$ \\
		
		\hline\hline
		
	\end{tabular}
\end{table*}

\subsubsection{Results Without II Signal}

We first set the data vector $C^{(1)}=C^{GG}+C^{IG}+C^{GI}+C^{GG,N}$ to investigate the efficiency of cleaning the $C^{IG}$ contamination by using SC.  Fig.~\ref{shift no II} shows the residual shifts of the SC method for LSST, when assuming II signals are negligible in all spectra with the binning method of SC (or can be cleaned by some other method). Two similar results with small shifts are obtained when applying our SC formalism to WFIRST and Euclid. Some numerical details are included in Table\,\ref{SC vs marg}, where the values are defined as fractional shifts in best-fit cosmological parameters:
\begin{equation} \label{Delta p}
\Delta p\equiv \frac{p^{\rm best-fit}-p^{\rm fid}}{p^{\rm fid}} .
\end{equation}

Under this condition of no II signals, the residual shifts are only affected by the residual IG signals. For non-adjacent bins ($i<j-1$) the residual $C^{IG}_{ij}$ is around $1\%$ to $8\%$ level for the IA model we use in the data vector. For adjacent bins ($i=j-1$), the residual $C^{IG}_{ij}$ is around $5\%$ to $15\%$ level. This agrees with Zhang 2008 \cite{SC.Zhang} that SC's efficiency is high when the 2 bins are far away and is relatively lower at close bins.  All these residuals lead to the $<1\sigma$ residual shift (of the black contours) in the best-fit cosmology in Fig.~\ref{shift no II} for LSST and similar small shifts for WFIRST and Euclid. \textcolor{black}{This gives the ideal performance of SC. But the binning method of SC is not perfect so the II signals will have some impact. The SC technique in this work minimizes the II signals but doesn't clean them. In future studies, other methods like Ref.\,\cite{Zhang2010}, which is another SC method, could potentially be used to clean the II contamination and bring the SC efficiency to the level introduced in this subsection. Next we are going to show how the contours will change when II signals are added.} 

\begin{table*}[t]
	\caption{\label{SC vs marg} The performance of SC with parameter forecasts for different surveys: LSST, WFIRST, and Euclid. $\Delta p$ presents the fractional residual bias of the best-fit cosmological parameters shifting from the fiducial values, as shown in Eq.\,\eqref{Delta p}. This table gives the numerical details for LSST (Fig.\,\ref{shift no II}), WFIRST and Euclid, assuming II signals are negligible with the binning method of SC. We can see that, for the 3 surveys, all the residual shifts are small and at the $1\%$ level. The residual shifts of LSST and WFIRST are more similar, mainly because their redshift distributions are more similar. The last column of $w_a$ is not in percentage form because its fiducial value is 0.}
		\begin{tabular}{  c | c | c | c | c | c | c  }
			\hline\hline
			\multirow{2}{*}{Survey} & \multirow{2}{*}{Method}
			& \multicolumn{5}{|c|}{$\Delta p$} \\
			\cline{3-7}
			& & $\Omega_m$ & $h_0$ & $\sigma_8$ & $w_0$ & $w_a$ \\
			\hline
			LSST & SC + marg photo-z & $0.85\%$ & $-1.56\%$ & $-0.25\%$ & $2.10\%$ & $-0.0565$ \\ 
			\cline{2-7}
			
			\hline
			WFIRST & SC + marg photo-z & $0.68\%$ & $-0.99\%$ & $-0.21\%$ & $1.08\%$ & $-0.0709$ \\ 
			\cline{2-7}
			
			\hline
			Euclid & SC + marg photo-z & $0.45\%$ & $-1.07\%$ & $0.04\%$ & $1.42\%$ & $-0.1976$ \\ 
			\cline{2-7}
			\hline\hline
			
		\end{tabular}
\end{table*}

\begin{table*}[t]
	\caption{\label{SC vs IA} The improvement in the best-fit cosmological parameters when SC is applied, with respect to the full IA contamination case. For the 3 surveys, the calculated values are within $\sim 10\%$ meaning that after applying the SC, the residual shifts of IA contamination is reduced to under  $\sim0.1$ level of the shifts of the full IA contamination. This agrees with the top-left panel of Fig.\,\ref{Reproduce Zhang} calculated as Eq.\,\ref{epsilon}. For the 3 different surveys, 2 different methods of calculation are applied: with II signals in the data vectors (the actual outcome of SC), or without II signals (the ideal case of SC with perfect binning method) in the data vectors. As shown in Fig.\,\ref{shift no II} and \ref{shift with II}, the II signals do make a difference in the directions of the shifts as well as the in the magnitudes.}
	\begin{tabular}{  c | p{0.1\textwidth}<{\centering} | p{0.1\textwidth}<{\centering} | p{0.1\textwidth}<{\centering} | p{0.1\textwidth}<{\centering} | p{0.1\textwidth}<{\centering} | p{0.1\textwidth}<{\centering}  }
		\hline\hline
		\multirow{2}{*}{Survey} & \multirow{2}{*}{Method}
		& \multicolumn{5}{|c|}{$\Delta p^{SC}/\Delta p^{IA}$} \\
		\cline{3-7}
		& & $\Omega_m$ & $h_0$ & $\sigma_8$ & $w_0$ & $w_a$ \\
		\hline
		\multirow{2}{*}{LSST}
		& with II & $5.75\%$ & $8.11\%$ & $5.57\%$ & $-0.94\%$ & $-0.08\%$ \\
		\cline{2-7}
		& without II & $-2.84\%$ & $-5.70\%$ & $-2.60\%$ & $3.61\%$ & $-2.04\%$ \\
		
		\hline
		\multirow{2}{*}{WFIRST}
		& with II & $9.14\%$ & $13.56\%$ & $9.21\%$ & $-2.13\%$ & $-1.81\%$ \\
		\cline{2-7}
		& without II & $-2.30\%$ & $-4.12\%$ & $-2.30\%$ & $1.92\%$ & $-2.95\%$ \\
		
		\hline
		\multirow{2}{*}{Euclid}
		& with II & $6.25\%$ & $14.49\%$ & $6.02\%$ & $0.33\%$ & $2.33\%$ \\
		\cline{2-7}
		& without II & $-1.41\%$ & $-6.38\%$ & $-0.43\%$ & $1.48\%$ & $-3.39\%$ \\
		
		\hline\hline
		
	\end{tabular}
\end{table*}

%

\subsubsection{Results With II Signal}

In SC, the intrinsic II signals are cleaned by the binning method. For non-adjacent bins, $C^{II}$ is significantly smaller than the residual $C^{IG}$ so the binning method is valid. However in the adjacent bins, $C^{II}$ is more dominant than the residual $C^{IG}$, and it causes a residual shift in the opposite direction in parameter space, see Fig.~\ref{shift with II} and Table III. The residual shifts are not significantly large, nonetheless, the shifts in $\Omega_m$, $h_0$ and $\sigma_8$ are larger with the II signal than the ones without the II signal, by approximately a factor of two. So to get better results with SC, dealing with II in the adjacent bins is still necessary. Moreover, cleaning II will also make SC applicable to the auto-spectra. In this way the constraining power in the auto-spectra will be available. If measurement of $C^{II}$ is applicable, it is expected to have not only smaller residual shifts in the best-fit values, but also significantly smaller contours than our current result shown in Fig.~\ref{shift with II}. This can also potentially affect the binning method of SC.

\textcolor{black}{We also compare the residual shifts from the SC to those from the marginalization from Ref.\cite{Krause2016}. Despite the differences between the SC and the marginalization methods, we find the two shifts are of the same order of magnitude. But the directions of the shifts can be similar or different depending on a number of factors, including the assumed underlying IA model, the IA model for the marginalization, and the efficiency of SC. And we do observe such similarities or discrepancies in the direction of the shifts. For example, the directions of the residual shifts without II generally agree with those of Ref.\,\cite{Krause2016}, while the shifts with II signals are in opposite directions. This is mainly because the II signals dominate the adjacent bins and lead to the opposite shifts.}

In Table.\,\ref{SC vs IA}, we showed the improvement in the best-fit cosmological parameters when SC is applied, comparing to the ones with the full IA contamination, i.e. not dealing with the IA problem at all. \textcolor{black}{The reason why we are not comparing the SC case with the marginalization case is because the shift for the marginalization case is very model-dependent, thus the comparison will be not very meaningful.} SC is expected to clean at least $\sim90\%$ of the IA signal, thus we expect the residual shift to be at $\sim0.1$ level of the full IA shift. $\Delta p^{SC}$ is the residual shift of SC, which is given by Eq.\,\eqref{Newtonian shift}, while $\Delta p^{IA}$ represents the shift of best-fit cosmological parameters with the full IA contamination, calculated by the same equation but uses full IA (IG+GI+II) spectra rather than the residual IA spectra in the partial derivative $A_\alpha$ as in Eq.\,\eqref{A_alpha}.

Some of the directions of our residual shifts are not in exact agreement with Ref.\,\cite{Krause2016}, for example $w_0$ and $w_a$. Such differences could be caused by the fact that the efficiency of SC differs for different $\ell$ values, as shown in the top-left panel of Fig.\,\ref{Reproduce Zhang}. The efficiency is generally lower at small $\ell$ and higher at large $\ell$. The dependencies of $C^{GG}$ spectra on the cosmological parameters also vary at different $\ell$ values. Thus the instability of how the efficiency of SC depends on $\ell$ will affect the directions of the residual shifts in cosmological parameters. Another possible reason is we are not including either the auto-spectra or the mitigation of II-spectra, which differs from their analysis. Also the residual shifts can be affected by the choice of tomographic bins. As we discussed before in the photo-z model subsection, in this analysis we are using only the major part of the survey with a constant bin-width, which differs from their binning method. The fact that they included the non-Gaussian parts of the covariance in their analysis, which our Fisher formalism doesn't include, is another possible reason. We therefore suspect these are the possible reasons for the difference between the estimated shifts of $w_0$ and $w_a$. This again emphasizes the importance of accurate $C^{II}_{ii}$ measurement, which we will discuss in a follow-up paper.

\section{Summary and conclusions} \label{Section discussion}

We developed and applied a formalism to estimate the effects of the self-calibration of the GI intrinsic alignment (IA) signal on the precision and accuracy of cosmological parameter constraints. We derived expressions for the self-calibration Fisher matrix Eq.\,\eqref{Fisher SC} and the spectra covariance Eq.\,\eqref{SC Cov} by propagating the involved errors of cosmic shear as well as the error of $C^{Ig}$ measurement Eq.\,\eqref{scaling-2}, shown in the Appendix.

We implemented three situations to compare the IA self-calibration application case to the case with no-systematics and then to the IA marginalization method case. Each mitigation method will improve the accuracy in determining the cosmological parameters from a lensing survey at the price of losing a little bit precision which is what we quantified in this analysis.

We find that the confidence contour increase and the accuracy gained by applying the self-calibration are very comparable to those obtained for the marginalization method. This indicates that the two methods are equally competitive although they operate very differently.

We analyzed the accuracy gained from applying the self-calibration by calculating the residual shift in the best fit cosmological parameters. In other words, we calculate the residual shift after the self-calibration method has eliminated most of the bias in the parameters due to the GI. As shown in Fig.\,\ref{SC vs IA}, by applying SC we can reduce the shift in cosmological parameters due to IA by at least one order of magnitude, for example the residual shift  in $\Omega_m$ is $5.75\%$ of the shift of the full IA contamination for LSST.

It is worth noting that a distinguishing characteristic of the self-calibration method is that it is not necessary to assume an IA model in the process. Once the IA signal is extracted, it can serve for studies to better understand structure formation scenarios and IA modeling. The two methods discussed are thus complementary from not only the point of view of IA mitigation, but also their modeling as well.

\textcolor{black}{It is worth pointing out that a degeneracy is present between the galaxy bias $b_i$ and $\sigma_8$ when applying the SC method. This degeneracy can lead to extra measurement uncertainty on $b_i$. To deal with it, one has to assume that we have good constraints on $C^{mm}$ from, for example from CMB observations, see for example Section 3.2 in Zhang 2008 \cite{SC.Zhang}. But for future lensing surveys, CMB alone may not be able to provide strong enough constraints to do that as for example for the late-time $C^{mm}$ in models with a time-dependent dark energy equation of state $w(a)$. To obtain such strong constraints, one will have to use combined probes such as CMB+SN+BAO. However, this is extra information that is needed by the SC compared to the marginalization method, even though SC doesn't require any statistical constraining power from these experiments.}

We performed our analyses for future cosmic shear surveys LSST, WFIRST and Euclid and found that the self-calibration offers a competitive and promising method to extract and mitigate the GI intrinsic alignment signal in order to allow these cosmic shear surveys to reach their full potential.

\begin{acknowledgements}

We thank M. Kesden, L. King, T. Kitching, P. Zhang and X. Zhao for useful discussions; J. Zuntz for useful suggestions and comments on using CosmoSIS (\url{https://bitbucket.org/joezuntz/cosmosis/wiki/Home}); and O. Dore, T. Kitching, J. Rhodes and T. Tyson  for providing survey specifications for LSST, Euclid and WFIRST. We also thank L. Fox for proofreading the manuscript. M.I. acknowledges that this material is based upon work supported in part by NSF under grant AST-1517768 and an award from the John Templeton Foundation.

\end{acknowledgements}

\bibliography{references}
\bibliographystyle{apsrev4-1}

\begin{widetext}
\appendix*
\section{Derivation of the Covariance}

According to the scaling relation Eq.~(\ref{scaling-1}), we can measure the IA spectrum $C^{IG}_{ij}$. The measurement uncertainty using Self Calibration $\Delta C^{IG}_{ij}$ can be expressed using the propagation of uncertainty:
\begin{align}
\Delta C^{IG(SC)}_{ij}&\approx\Delta\left(\frac{W_{ij}(l)\Delta_i(l)}{b_i(l)}C^{Ig}_{ii}(l)\right)\notag\\
&=\sqrt{(\frac{W_{ij}\Delta_i}{b_i})^2(\Delta C^{Ig}_{ii})^2+(-\frac{W_{ij}\Delta_iC^{Ig}_{ii}}{b_i^2})^2(\Delta b_i})^2\notag\\
&=C^{IG(SC)}_{ij}\sqrt{(\frac{\Delta C^{Ig}_{ii}}{C^{Ig}_{ii}})^2+(\frac{\Delta b_i}{b_i})^2}.
\end{align}

As $W_{ij}$ and $\Delta_i$ are theoretical values, there are no measurement errors. $\Delta C^{Ig}_{ii}$ is the measurement uncertainty that comes from the introduced observable $C^{(2)}$. $\Delta b_i$ is the measurement uncertainty that comes from observable $C^{(3)}$. The 2 uncertainties derived in Zhang's SC paper \cite{SC.Zhang} are reproduced here:
\begin{align}
 \label{shot error 1}
(\Delta C^{Ig}_{ii})^2&=\frac{1}{2l\Delta l f_{\rm sky}} \left( {C^{gg}_{ii}C^{GG}_{ii}+\left[1+\frac{1}{3(1-Q)^2}\right]}\right.\notag\\
&\times[C^{gg}_{ii}C^{GG,N}_{ii}+C^{gg,N}_{ii}(C^{GG}_{ii}+C^{II}_{ii})] \notag\\
&\left.{+C^{gg,N}_{ii}C^{GG,N}_{ii}\left[1+\frac{1}{(1-Q)^2}\right]}  \right),\\
 \label{shot error 2}
\frac{\Delta b_i}{b_i}&\sim\frac{1}{2}\sqrt{\frac{1}{l\Delta lf_{\rm sky}}}\times\left(1+\frac{C^{gg,N}_{ii}}{C^{gg}_{ii}}\right),
\end{align}
in which $C^{gg,N}_{ii}=4\pi f_{\rm sky}/N_i$, $C^{GG,N}_{ii}=4\pi f_{\rm sky}\gamma_{\rm rms}^2/N_i$. \textcolor{black}{The noise spectrum in this work is assumed to be perfectly known thus there is no need to marginalize over it. In Eq.\,\eqref{shot error 2}, the error from $C^{mm}$ according to Eq.\,\eqref{data measured b_i} is assumed to be negligible according to Ref.\,\cite{SC.Zhang}. For the same reason, as $C^{mm}$ is tightly constrained, the degeneracy between $\sigma_8$ and galaxy bias (which will lead to some extra measurement error) is assumed to be negligible in this work.} ``N'' stands for measurement noise, $N_i$ is the total number of galaxies in i-th redshift bin. The multiple bin size $\Delta l=0.2l$ for LSST.

According to Zhang, Eq.~(\ref{shot error 2}) is an approximated value, as $C^{gg}_{ii}\approx b_i^2 C^{mm}_{ii}$, and $C^{mm}_{ii}$ is the matter angular power spectrum, which is tightly constrained. Hence $C^{mm}_{ii}$ can be considered as a theoretical value without any measurement error, comparing with the measurement error on $C^{(3)}$.

By propagating the measurement noise from $C^{(2)}$ and $C^{(3)}$ to $C^{IG(SC)}$, we have shown that, as in Fig.~\ref{statistical error}, the contribution from $C^{(3)}$ (red) is much smaller than that of $C^{(2)}$ (blue). Thus in this analysis we will focus on the extra noise from $C^{(2)}$.

We therefore derive the values in Eq.~(\ref{SC Cov}).
\begin{align}
&Cov(C^{GG(SC)}_{ij},C^{GG(SC)}_{pq})=Cov\left(C^{(1)}_{ij}-C^{IG(SC)}_{ij},C^{(1)}_{pq}-C^{IG(SC)}_{pq}\right) \\\notag
&=Cov(C^{(1)}_{ij},C^{(1)}_{pq})-Cov(C^{(1)}_{ij},{C}^{IG(SC)}_{pq})-Cov({C}^{IG(SC)}_{ij},C^{(1)}_{pq})+Cov(C^{IG(SC)}_{ij},C^{IG(SC)}_{pq}).
\end{align}

Within the 4 different covariance terms above, the 1st is well known as the covariance of the observed cosmic shear spectra.

\begin{align}
&Cov(C^{(1)}_{ij},C^{(1)}_{pq})=\frac{1}{2l+1}(C^{(1)}_{ip}C^{(1)}_{jq}+C^{(1)}_{iq}C^{(1)}_{jp}) \label{Cov_term1}.
\end{align}

The 2nd term is calculated here as:
\begin{align}
Cov(C^{(1)}_{ij},{C}^{IG(SC)}_{pq})&=<\left(C^{(1)}_{ij}-<C^{(1)}_{ij}>\right)\left({C}^{IG(SC)}_{pq}-<{C}^{IG(SC)}_{pq}>\right)> \\\notag
&=<C^{(1)}_{ij}{C}^{IG(SC)}_{pq}>-<C^{(1)}_{ij}><{C}^{IG(SC)}_{pq}>.
\label{Cov_term2}
\end{align}
by definition, where $<...>$ denotes the ensemble average.

For the given photo-z bins $\{ i,j,p,q \}$, we pixelate the data such that pixel $\alpha$ is within the i-th bin $\alpha\in i$, with number of pixels $N_P$, measured overdensity $(\delta+\delta^N)^m_\alpha$ and shear $(\gamma^G+\gamma^N)^m_\alpha$ with $-l\leq m\leq l$. We define $(\gamma^G+\gamma^N)^m_\alpha \equiv (\gamma^G+\gamma^N)e^{i\vec{l}\cdot\vec{\theta_\alpha}}$, in which $\vec{l}=(l,m)$. Similarly we can have pixels $\{\beta,\rho,\sigma\}$ that obey $\beta\in j$, $\rho\in p$, $\sigma\in p$. Thus, by using the definition of Eq.~(\ref{scaling-1}) and Eq.~(\ref{scaling-2}) we have,
\begin{align}
<\hat{C}^{(1)}_{ij}C^{IG(SC)}_{pq}>=&\frac{W_{pq}\Delta_p}{b_p}\frac{1}{1-Q_p}\frac{1}{(2l+1)^2}\sum\limits_{m,m'}N_P^{-4} \sum\limits_{\alpha\beta\rho\sigma} \\
&<(\gamma^G+\gamma^I+\gamma^N)^m_\alpha
(\gamma^G+\gamma^I+\gamma^N)^{-m}_\beta
(\delta+\delta^N)^{m'}_\rho
(\gamma^G+\gamma^I+\gamma^N)^{-m'}_\sigma
(2S_{\rho\sigma}-Q)>. \notag
\end{align}
in which $S_{\rho\sigma}$ is either 1 (for photo-z $z^P_\rho>z^P_\sigma$) or 0 (for $z^P_\rho \leq z^P_\sigma$). For a large enough $N_P$, $\sum\limits_{\rho\sigma}S_{\rho\sigma}=N_P^2/2$ so that the average is $\bar{S}_{\rho\sigma}=1/2$.

For Gaussian fields \{A,B,C,D\}, the 4-point correlation can be expanded according to Wick's theorem:
\begin{equation}
<ABCD>=<AB><CD>+<AC><BD>+<AD><BC>.
\end{equation}

Therefore the above ensemble average equation can be expressed as
\begin{align}
<\hat{C}^{(1)}_{ij}C^{IG(SC)}_{pq}>=&\frac{W_{pq}\Delta_p}{b_p}\{C^{(1)}_{ij}C^{Ig}_{pp}+\frac{1}{2l+1}\left[(C^{Gg}_{ip}+C^{Ig}_{ip})C^{(1)}_{jp}+C^{(1)}_{ip}(C^{Gg}_{jp}+C^{Ig}_{jp})\right]\},
\end{align}
based on the definition of $Q_i={C^{Gg}_{ii}|_S}/{C^{Gg}_{ii}}$ , $C^{Gg}_{pp}$ term vanishes.

Thus the final expression for Eq.~(\ref{Cov_term2}) is
\begin{align}
Cov(C^{(1)}_{ij},C^{IG(SC)}_{pq})&=\frac{1}{2l+1}\frac{W_{pq}\Delta_p}{b_p}\left[(C^{Gg}_{ip}+C^{Ig}_{ip})C^{(1)}_{jp}
+C^{(1)}_{ip}(C^{Gg}_{jp}+C^{Ig}_{jp})\right] \label{Cov2}.
\end{align}.

By symmetry, the 3rd covariance terms will be,

\begin{align}
Cov(C^{IG(SC)}_{ij},C^{(1)}_{pq})&=\frac{1}{2l+1}\frac{W_{ij}\Delta_i}{b_i}\left[(C^{Gg}_{pi}+C^{Ig}_{pi})C^{(1)}_{qi}
+C^{(1)}_{pi}(C^{Gg}_{qi}+C^{Ig}_{qi})\right] \label{Cov3}.
\end{align}

Similarly, the 4th covariance term

\begin{align}
\label{Cov4}
Cov( C^{IG(SC)}_{ij}, C^{IG(SC)}_{pq} ) \approx&  \frac{1}{2l+1}\frac{W_{pq}\Delta_p}{b_p}\frac{W_{ij}\Delta_i}{b_i} \{ C^{gg}_{ip}C^{(1)}_{ip} + \delta_{ip}C^{ggN}_{ii}C^{(1)}_{ip}[1+\frac{1}{3(1-Q_i)^2}] + C^{gg}_{ip}\delta_{ip}C^{GGN}_{ii}[1+\frac{1}{3(1-Q_i)^2}] \\
&+ \delta_{ip}C^{ggN}_{ii}C^{GGN}_{ii}[1+\frac{1}{(1-Q_i)^2}] +(C^{gG}_{ip}+C^{gI}_{ip})(C^{Gg}_{ip}+C^{Ig}_{ip}) \}. \notag
\end{align}.

Here because noise can only correlate at zero-lag (this will also result in the Kronecker delta), the following relations are applied to the above covariance:
\begin{subequations}
\begin{align}
	\frac{1}{N_P^4}\sum\limits_{\alpha\beta\rho\sigma}(2S_{\alpha\beta}-Q)(2S_{\rho\sigma}-Q)&\approx (1-Q)^2, \\
	\frac{1}{N_P^3}\sum\limits_{\alpha\beta\rho}(2S_{\alpha\beta}-Q)(2S_{\rho\beta}-Q)&\approx (1-Q)^2 + \frac{1}{3}, \\
	\frac{1}{N_P^2}\sum\limits_{\alpha\beta}(2S_{\alpha\beta}-Q)^2&\approx (1-Q)^2 +1.
\end{align}
\end{subequations}

The 4th covariance agrees with the Zhang 2008 SC paper \cite{SC.Zhang}. By combining the expressions in Eq.~(\ref{Cov_term1}), (\ref{Cov2}), (\ref{Cov3}) and (\ref{Cov4}) we can get the final expression for Eq.~(\ref{SC Cov}).

Fig.\,\ref{contour}, \ref{shift no II} and \ref{shift with II} are results using the full expression of Eq.\,\eqref{SC Cov}. Because of the limit of computational accuracy, the inverse of large matrices can not be taken properly. We use the function $pinv$ (which presents a Moore-Penrose pseudoinverse to the matrix) in Matlab (also exists in GNU, Octave or Numpy) to numerically solve this problem. Another alternative solution is to use the singular value decomposition, which is given by the Fourier transformation. We are developing a CosmoSIS \cite{CosmoSIS} module for the calculation of SC terms. However, we found that even the $pinv$ function has some instabilities, as the covariance of Eq.\,\eqref{SC Cov} is very large ($10^2\times 10^2$ in our case). When the instabilities happen, the shift changes significantly. We therefore applied the following approximation to compare with the results we presented in this paper. The comparison of the plots shown in the main paper have good agreement, thus we believe $pinv$ does give stable calculations of our results.

To get a more stable estimation of parameter shift, we adopt the approximation as follows:
\begin{equation}\label{Cov_approx}
Cov(C^{GG(SC)}_{ij},C^{GG(SC)}_{pq})\approx Cov(C^{GG}_{ij},C^{GG}_{pq}).
\end{equation}
In this way it will simply become the covariance of the cosmic shear spectra. The resulting Fisher matrix of this approximation is $\sim14\%$ smaller than using the full expression of Eq.\,\eqref{SC Cov}, and the associated Covariance matrix is $~12\%$ greater, meaning that the error is enlarged by a $\sim6\%$ level. This agrees with what has been shown in Ref.\,\cite{SC.Zhang} and our Fig.\,\ref{statistical error}: using SC doesn't introduce much change in the uncertainties. And the pseudo-inverse of the covariance of Eq.\,\eqref{Cov_approx} is much more stable.

We will also show the shift of best-fit cosmological parameters using this approximation of Eq.\,\eqref{Cov_approx}. Fig.\,\ref{shift no II approx} using Eq.\,\eqref{Cov_approx} and Fig.\,\ref{shift no II} using Eq.\,\eqref{SC Cov} do have good agreement. Therefore this approximation is applicable for future studies.

\begin{figure*}[h]
	\includegraphics[width=0.9\columnwidth]{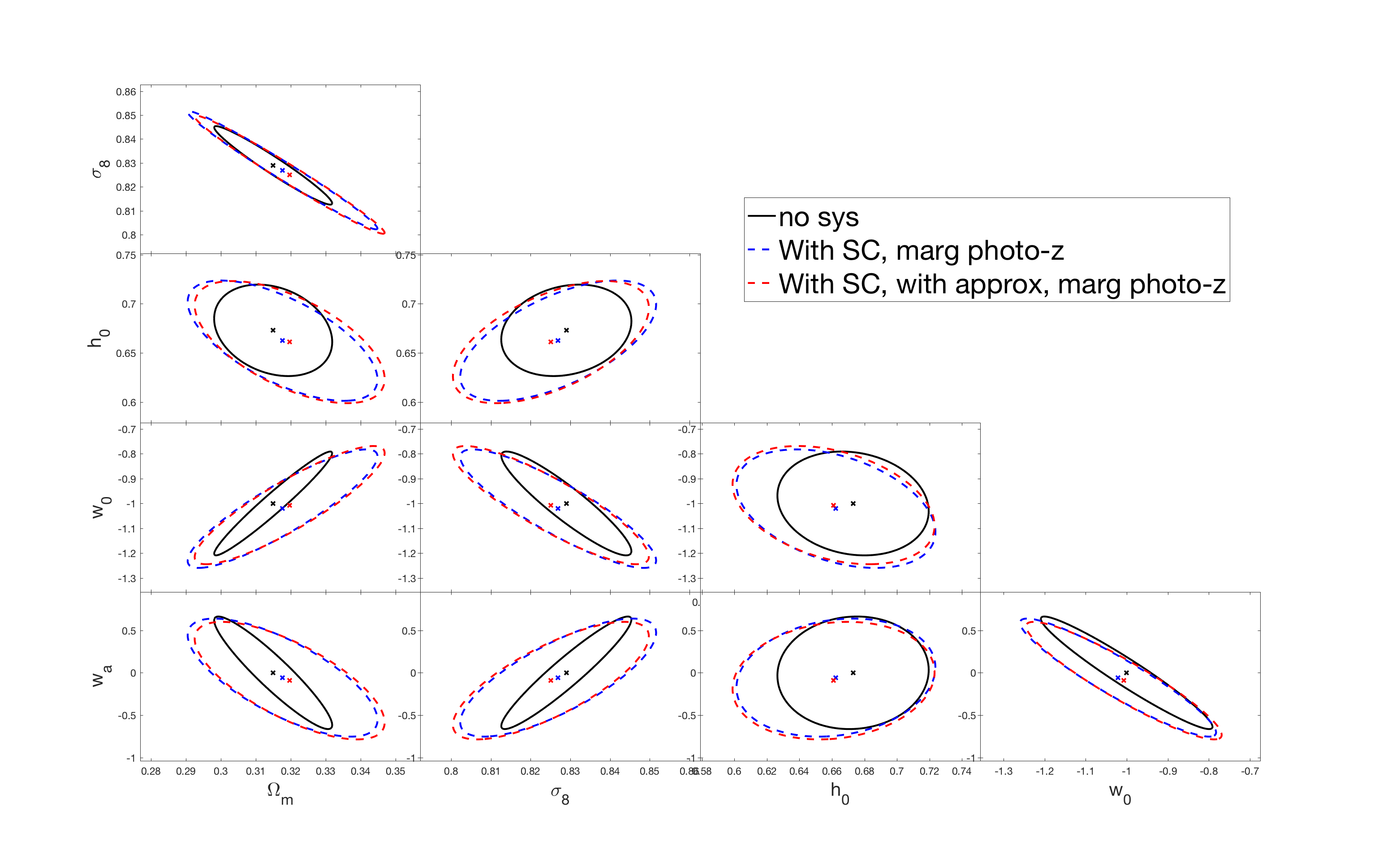}
	\caption{Shift in best-fit cosmology when the approximation Eq.\,\eqref{Cov_approx} is applied. Blue contours are exactly the same as in Fig.\,\ref{shift no II} when no approximation is applied, while red contour uses the approximation introduced in Eq.\,\ref{Cov_approx}. The blue and red have good agreement which shows that the pseudoinverse is stable, and confirms that SC doesn't introduce much change in the uncertainties. So that in the future we can use this approximation to give a fast result without deriving the complicated expression of the covariance. We also tested intentionally increasing the residual IA signal, leading to larger shifts in both blue and red. The shifts are still in very good agreement.
		\label{shift no II approx}}
\end{figure*}

\end{widetext}

\end{document}